\documentclass[aps,showpacs]{revtex4}
\usepackage{amsmath}
\usepackage{amssymb}
\usepackage{graphicx}% Include figure files
\usepackage{dcolumn}% Align table columns on decimal point
\usepackage{bm}% bold math

\newcommand{\la}{\lambda}
\newcommand{\ga}{\gamma}
\newcommand{\al}{\alpha}

\newcommand{\br}{{\bf r}}
\newcommand{\prt}{\partial}
\newcommand{\bu}{{\bf u}}

\begin{document}

\title{
The theory of optical dispersive shock waves in photorefractive media}
\author{G.A. El$^{1}$}
\email{G.El@lboro.ac.uk}
\author{A. Gammal$^{2}$}
\email{gammal@if.usp.br}
\author{E.G. Khamis$^{2}$}
\email{egkhamis@if.usp.br}
\author{R.A. Kraenkel$^{3}$}
\email{kraenkel@ift.unesp.br}
\author{A.M. Kamchatnov$^{4}$}
\email{kamch@isan.troitsk.ru}

\affiliation{ $^1$ Department of Mathematical Sciences, Loughborough
University, Loughborough LE11 3TU, UK \\
$^2$ Instituto de F\'{\i}sica, Universidade de S\~{a}o Paulo,
05315-970, C.P.66318 S\~{a}o Paulo, Brazil\\
$^3$Instituto de F{i}sica Te\'{o}rica, Universidade Estadual
Paulista, Rua Pamplona 145, 01405-900 S\~{a}o Paulo, Brazil\\
$^4$Institute of Spectroscopy, Russian Academy of Sciences, Troitsk,
Moscow Region, 142190, Russia\\ }

\date{\today}

\begin{abstract}
The theory of optical dispersive shocks generated in propagation of
light beams through photorefractive media is developed. Full
one-dimensional analytical theory based on the Whitham modulation
approach is given for the simplest case of sharp step-like initial
discontinuity in a beam with one-dimensional strip-like geometry.
This approach is confirmed by numerical simulations which are
extended also to beams with cylindrical symmetry. The theory
explains recent experiments where such dispersive shock waves have
been observed.
\end{abstract}


\maketitle

\section{Introduction}

Study of optical solitons is a large area of modern research which
is important both scientifically and for potential applications
(see, e.g., \cite{ka03,kld}). Different kinds of solitons have
already been observed in various nonlinear optical media and their
behavior has been explained in the frameworks of such mathematical
models as nonlinear Schr\"odinger (NLS) and generalized nonlinear
Schr\"odinger (GNLS) equations for different dimensions and
geometries, so that one can consider the properties of single
solitons as well enough understood.

However, there are situations when many solitons are generated so
that they comprise a dense soliton lattice. In such situations, it
is impossible to neglect interactions between solitons and one has
to consider evolution of the structure as a whole rather than to
trace evolution of each soliton separately. Usually, such soliton
structures appear as a result of wave breaking of a large enough
initial pulse or large disturbance along constant background. Hence,
such structures  can be considered as a dispersive counterparts of
shock waves well known in physics of compressible viscous fluids
(see, e.g., \cite{whitham-1}). In a viscous fluid, the shock can be
represented as a narrow region within which strong dissipation
processes take place. In optics, on the contrary, dissipation
effects can be neglected compared with dispersion ones and the shock
discontinuity resolves into an expanding region filled with
nonlinear oscillations. Such dispersive shock waves are known as
tidal bores in rivers \cite{bl54} and have been also observed in
some other physical systems including collisionless plasma
\cite{plasma} and Bose-Einstein condensate \cite{hoefer}.
Experiments on such dispersive shock waves production in optics have
been recently reported in \cite{french,fleischer}. Motivated by
these experiments, we shall consider here the theory of dispersive
shock waves in photorefractive media.

Since the number of interacting solitons in dispersive shocks is
usually much greater than unity and these solitons are spatially
ranked in amplitude, such a dispersive shock  can be represented as
a modulated periodic wave with parameters changing a little in one
transverse or longitudinal period of envelope amplitude of the
electromagnetic wave. A slow change of the parameters of the
envelope amplitude is governed to leading order by the Whitham
modulation equations obtained by averaging conservation laws over
the family of nonlinear periodic solutions or by the application of
the averaged variational principle (see, e.g.,
\cite{whitham-1,ri,kamch2000}). For the one-dimensional NLS
equation, the Whitham equations were derived in \cite{forest,pavlov}
(see also \cite{kamch2000}) and the mathematical theory of
dispersive shock waves for the defocusing case was developed in
\cite{gk87,eggk95, ek95, jin,tian,kku, bk}. It was applied to the
propagation of signals in optical fibers in \cite{kodama} and in
Bose-Einstein condensates in \cite{kgk04,hoefer}. It should be
mentioned that for the case of the 1D NLS equation, the presence of
an integrable structure has important consequences for the
modulation (Whitham) system, namely, the latter can be represented
in a diagonal (Riemann) form, which dramatically simplifies further
analysis. The method of obtaining the Whitham equations in this form
is based on the Inverse Scattering Transform (IST) applied to the
NLS equation \cite{forest},\cite{pavlov}. However, in the case of
the GNLS equation, the IST method cannot be used anymore, and the
diagonal structure of the Whitham system is not available.
Nevertheless, it was shown in \cite{el}-\cite{ekt} that in this
case, the main characteristics of the dispersive shock wave still
can be found by using some general properties of the Whitham
equations which remain present even in non-integrable case. Here we
shall use this latter method for derivation of parameters of
one-dimensional dispersive shock waves generated in photorefractive
crystals and shall confirm our analytical results by numerical
simulations, which also provide a more detailed information in the
cases when the analytical approach is not yet developed (say in 2D).

\section{Main equations}

Photorefractive optical solitons were first observed in the
experiment \cite{duree} and in
the experiments \cite{french,fleischer} the formation of
dispersive shock waves has been observed in spatial evolution of
light beams propagating through self-defocusing photorefractive
crystals, so that beam non-uniformities give rise to breaking
singularities and their resolution through  dispersive shocks. As is
known, propagation of such stationary beams is described by the
equation
\begin{equation}\label{1-1}
    i\frac{\prt\psi}{\prt z}+\frac1{2k_0}\Delta_\bot\psi
    +\frac{k_0}{n_0}
    \delta n\left(|\psi|^2\right)\psi=0,
\end{equation}
where $\psi$ is envelope field strength of electromagnetic wave
with wave number $k_0=2\pi n_0/\la$, $z$ is the coordinate
along the beam, $x,y$ are transverse coordinates, $\br =(x,y)$,
$\Delta_\bot=\prt^2/\prt^2x+\prt^2/\prt^2y$ is transverse Laplacian,
$n_0$ is a linear refractive index,
and in photo-refractive medium we have
\begin{equation}\label{1-2}
    \delta n=-\frac12 n_0^3 r_{33}E_p\frac{\rho}{\rho+\rho_d},
\end{equation}
where $E_p$ is applied electric field, $r_{33}$ electro-optical index,
$\rho=|\psi|^2$, and $\rho_d$ is a saturation parameter.

For mathematical convenience, we introduce non-dimensional variables
\begin{equation}\label{1-4}
    \tilde{z}=\frac12 kn_0^2r_{33}E_p\left(\frac{\rho_c}{\rho_d}\right) z,\quad
    \tilde{x}=kn_0\sqrt{\frac12r_{33}E_p\left(\frac{\rho_c}{\rho_d}\right)}x,\quad
    \tilde{y}=kn_0\sqrt{\frac12r_{33}E_p\left(\frac{\rho_c}{\rho_d}\right)}y,\quad
    \tilde{\psi}=\sqrt{\rho_c}\psi,
\end{equation}
where $\rho_c$ is a characteristic value of optical intensity (its concrete
definition depends on the problem under consideration; for instance, it
can the background intensity),
so that Eq.~(\ref{1-1}) takes the form of GNLS equation
\begin{equation}\label{1-5}
    i\frac{\prt\psi}{\prt z}+\frac1{2}\Delta_\bot\psi-
    \frac{|\psi|^2}{1+\gamma|\psi|^2}\psi=0,
\end{equation}
where $\gamma=\rho_c/\rho_d$ and tildes are omitted. If saturation
effect is negligibly small ($\gamma|\psi|^2\ll1$), then this
equation reduces to the usual NLS equation
\begin{equation}\label{1-6}
    i\frac{\prt\psi}{\prt z}+\frac1{2}\Delta_\bot\psi-
    {|\psi|^2}\psi=0.
\end{equation}

It is convenient to represent these equations in a fluid dynamics
type form by means of the substitution
\begin{equation}\label{1-7}
    \psi(\br, z)=\sqrt{\rho}\,\exp\left(i \int^{\br}
    \bu(\br´,z)d\br´\right) ,
\end{equation}
so that they are transformed to
\begin{equation}\label{1-8}
\begin{split}
\rho_z+\nabla_\bot(\rho \bu)=0, \\
\bu_z+\left(\bu\nabla_\bot\right)\bu +\nabla_\bot f(\rho)-\nabla_\bot
\left[\frac{\Delta_\bot\rho}{4\rho}-\frac{(\nabla_\bot\rho)^2}{8\rho^2}
\right] =0,
\end{split}
\end{equation}
where
\begin{equation}\label{2-1}
    f(\rho)=\frac{\rho}{1+\gamma\rho}\qquad \textrm{ for the GNLS equation (\ref{1-5})}
\end{equation}
and
\begin{equation}\label{2-2}
    f(\rho)={\rho}\qquad \textrm{ for the NLS equation
    (\ref{1-6})}\, .
\end{equation}
The light intensity $\rho$ in the hydrodynamic interpretation has a
meaning of a density of a ``fluid'' and Eqs.~(\ref{2-1}),
(\ref{2-2}) can be viewed as ``equations of state'' for such a
fluid. The function $\bu(\br,z)$ is a local value of the wave vector
component transverse to the direction of the light beam; in
hydrodynamic representation it has a meaning of the ``flow
velocity''. The variable $z$ plays the role of time so it is natural
to describe the deformations of the light beam in evolutionary
terms. Obviously, if initial distribution does not depend on one of
the transverse coordinates (say $y$), then transverse differential
vector operators reduce to usual derivatives ($\nabla_\bot=\prt/\prt
x$, $\Delta_\bot=\prt^2/\prt x^2$).

Evolution, according to (\ref{1-8}) of an initial distribution,
specified at $z=0$, typically leads to wave breaking and formation
of dispersive shock waves. One can distinguish the following typical
cases:

\begin{itemize}
\item{generation of dispersive shocks in the evolution of a bright strip hump
above a uniform (background) intensity distribution,}
\item{generation of sequences of solitons from a strip ``hole'' in the
light intensity},
\item{generation of dispersive shocks in the evolution of a bright cylindrically
symmetrical hump above a uniform intensity distribution}.
\end{itemize}
In 1D geometry such humps can be modeled qualitatively by step-like
pulses with sharp boundaries, and these models are convenient for
analytical considerations. As was shown in \cite{hoefer} for the NLS
equation case with $\ga=0$, this model agrees quite well with
numerical simulations of 2D dynamics. Therefore we shall start here
with these idealized models.

\section{Analytical theory of one-dimensional dispersive shocks
generated in decay of a step-like initial distribution}

We shall start with analytical treatment of shocks described by 1D
equation
\begin{equation}\label{3-1}
    i\psi_z+\frac12\psi_{xx}-f(|\psi|^2)\psi=0,
\end{equation}
or, in a fluid dynamics form, by the system
\begin{equation}\label{dh}
\begin{split}
\rho_z+(\rho u)_x=0, \\
u_z+uu_x+\frac{df}{d\rho}\rho_x+\left(\frac{(\rho_x)^2}{8\rho^2}
-\frac{\rho_{xx}}{4\rho} \right)_x=0,
\end{split}
\end{equation}
where the nonlinear refraction function $f(\rho)$ is given by
Eq.~(\ref{2-1}) or (\ref{2-2}). The systems of the type (\ref{dh})
are often referred to as dispersive hydrodynamics systems.

We consider initial distributions of the intensity and transverse
wave vector in the form
\begin{equation}\label{step}
\rho(x,0) =\left\{
\begin{array}{ll}
 \rho_0 &\quad \hbox{for} \quad x <0,\\
 1 & \quad \hbox{for} \quad x\ge0;
\end{array}
\right.
\qquad  u(x,0)=0,
\end{equation}
that is we assume that the initial velocity $u(x,0)$ is equal to
zero everywhere which means that the initial beam enters  the
photo-refractive medium at $z=0$ without any focusing. For the sake
of definiteness we assume also that  $\rho_0>1$.

At the initial stage of evolution,  linear waves are generated which
propagate according to the dispersion law obtained by means of
linearization of Eqs.~(\ref{dh}) about the uniform state
$\rho=\rho_0$, $u=u_0$ (we keep here a nonzero value of $u_0$ for
future convenience), that is $\rho=\rho_0+\rho_1\exp[i(kx-\omega
z)]$, $u=u_0+u_1\exp[i(kx-\omega z)]$, where $\rho_1,\, u_1\ll1$.
Then a simple calculation yields
\begin{equation}\label{dr}
\omega = \omega_0(\rho_0, u_0,k)=ku_0 \pm k
\sqrt{\frac{\rho_0}{(1+\gamma \rho_0)^2} +\frac{k^2}{4}} \, .
\end{equation}
Note that $\omega''(k)>0$, which implies appearance of {\it dark}
solitons in full nonlinear solutions. But before consideration of
such solutions, we shall discuss a nonlinear stage of evolution in
dispersionless approximation when one can neglect the higher order
terms in  the system (\ref{dh}). While in the case of general smooth
initial data this stage of evolution is responsible for the
formation of breaking singularities in the solution, its
consideration also provides important insights into the nonlinear
dissipationless dispersive dynamics of discontinuous disturbances of
the type (\ref{step}) even beyond the breaking point.

\subsection{Dispersionless approximation}

In dispersionless approximation, the system (\ref{dh}) reduces to
standard equations of compressible fluid dynamics
\begin{equation}\label{4-1}
\begin{split}
\rho_z+(\rho u)_x=0, \\
u_z+uu_x+{f'(\rho)}\rho_x=0.
\end{split}
\end{equation}
Because of the bi-directional nature of this system, generally, an
initial step (\ref{step}) resolves into a combination of two waves
propagating in opposite directions. One of these waves represents a
rarefaction wave with clear physical meaning, but the other one
leads to a multi-valued dependence of the intensity $\rho(x,z)$ and
transverse wave number (associated flow velocity) $u(x,z)$ on the
$x$--coordinate. Nevertheless, this formal global solution sheds
some light on the structure of the actual physical solution and some
its elements will be used later, therefore we shall consider it
here. To this end we cast the system (\ref{4-1}) into a diagonal
form (see, for instance, \cite{whitham-1,kamch2000}) by introduction
of new variables---Riemann invariants
\begin{equation}\label{4-2}
    r_{\pm}=u \pm \frac{2}{\sqrt{\gamma}}\arctan \sqrt{\gamma \rho},
\end{equation}
so that it takes the form
\begin{equation}\label{4-3}
\frac{\partial r_{\pm}}{\partial z} + V_{\pm} \frac{\partial
r_{\pm}}{\partial x} =0,
\end{equation}
where the characteristic
velocities $V_{\pm}$ are expressed in terms of the hydrodynamic
variables $\rho, u$ by the relationships
\begin{equation}\label{4-4}
\qquad V_{\pm}=u \pm \frac{\sqrt{\rho}}{1+\gamma \rho} .
\end{equation}
When $\gamma \to 0$ we have $r_{\pm}=u \pm 2\sqrt{\rho}$, $V_{\pm}=u
\pm \sqrt{\rho}$, i.e. the usual expressions for the dispersionless
limit of the defocusing NLS equation (the shallow-water
system---see, for instance, \cite{gk87}).

Since in the case of the step-like initial conditions the variables
$r_{\pm}$ must depend on a self-similar variable $\zeta=x/z$ alone,
the equations Eqs.~(\ref{4-3}) reduce to
$(V_{\pm}-\zeta)(dr_{\pm}/d\zeta)=0$ and we arrive at the so-called
simple-wave solutions:
\begin{equation}\label{4-5}
u + \frac{\sqrt{\rho}}{1+\gamma \rho}=\frac{x}z,\qquad u
-\frac{2}{\sqrt{\gamma}}\arctan \sqrt{\gamma
\rho}=r_-^0=\mathrm{constant},
\end{equation}
or
\begin{equation}\label{4-6}
u - \frac{\sqrt{\rho}}{1+\gamma \rho}=\frac{x}z,\qquad u +
\frac{2}{\sqrt{\gamma}}\arctan \sqrt{\gamma
\rho}=r^0_+=\mathrm{constant}.
\end{equation}
The constants here are chosen from the continuity conditions at the
points where the simple waves enter the regions of constant
intensities. Since the left-propagating rarefaction wave described
by (\ref{4-6}) matches with the external flow $\rho=\rho_0$, $u=0$
(see Fig.~1a) we have $ r_+^0=\frac{2}{\sqrt{\gamma}}\arctan
\sqrt{\gamma \rho_0}$ and, correspondingly
\begin{equation}\label{4-9}
    u=\frac2{\sqrt{\gamma}}\left({\arctan\sqrt{\gamma\rho_0}}-
    \arctan{\sqrt{\gamma\rho}}\right).
\end{equation}
Now, substituting this into the first equation (\ref{4-6}) we get
\begin{equation}\label{4-8}
    \frac{\sqrt{\rho}}{1+\gamma \rho}+\frac2{\sqrt{\gamma}}
    \left(\arctan{\sqrt{\gamma\rho}}-
    \arctan{\sqrt{\gamma\rho_0}}\right)=-\frac{x}z \, ,
\end{equation}
which determines implicitly the intensity $\rho$ as a function of
$x/z$ in the rarefaction wave. For $x <x_1^-$ we have $\rho=\rho_0 =
\mathrm{constant}$ so $x=x_1^-$ is the point of weak discontinuity
which must propagate with sound velocity (see, for instance
\cite{lanlif}) which in our case is
\begin{equation}\label{4-7}
    c_s(\rho)= \frac{\sqrt{\rho}}{1+\gamma \rho},
\end{equation}
Indeed, substituting $\rho = \rho_0$ into (\ref{4-8}) we get
$x^-_1/z=-c_s(\rho_0)$. As a matter of fact, the speeds of
propagation of weak discontinuities in the photorefractive system
agree with the group speeds determined by the long wavelength limit
$k\to 0$ in the linear dispersion relation (\ref{dr}).

Next, for $x>x_2^-$ we have $\rho=1$, $u=0$ (see Fig.~1a) and this
does not agree with the relationship (\ref{4-9}) in the constructed
left-propagating rarefaction wave solution. Hence, we have to
introduce some intermediate distribution
\begin{equation}\label{5-1}
    \rho(x/z)=\rho^-=\mathrm{constant},\quad
    u(x/z)=u^-=\mathrm{constant}
\end{equation}
which matches with the rarefaction wave at some $x=x_1^+$. Now, to
connect the intermediate distribution (\ref{5-1}) with $\rho=1$,
$u=0$ downstream, we have to use the right-propagating simple wave
solution (\ref{4-5}) where the constant  $r^0_+ =
\frac{2}{\sqrt{\gamma}}\arctan \sqrt{\gamma}$. Hence we get
\begin{equation}\label{5-3}
    u=\frac2{\sqrt{\gamma}}\left({\arctan\sqrt{\gamma\rho}}-
    \arctan{\sqrt{\gamma}}\right)
\end{equation}
and
\begin{equation}\label{5-2}
    \frac{\sqrt{\rho}}{1+\gamma \rho}+\frac2{\sqrt{\gamma}}
    \left({\arctan\sqrt{\gamma}}-
    \arctan{\sqrt{\gamma\rho}}\right)=\frac{x}z \, .
\end{equation}

Equations (\ref{4-9}) and (\ref{5-3}) at $\rho=\rho^-$ must give
$u=u^-$; hence they yield the equation
\begin{equation}\label{5-4}
    \arctan\sqrt{\gamma\rho^-}=\frac12\left(\arctan\sqrt{\gamma\rho_0}+
    \arctan\sqrt{\gamma}\right)
\end{equation}
which determines the parameter $\rho^-$:
\begin{equation}\label{5-5}
    \rho^-=\left[\frac{\sqrt{1+\gamma\rho_0}-1+\sqrt{\rho_0}(\sqrt{1+\ga}-1)}
    {\ga\sqrt{\rho_0}-(\sqrt{1+\ga\rho_0}-1)(\sqrt{1+\ga}-1)}\right]^2.
\end{equation}
When $\rho^-$ is  known, the parameter $u^-$ is found from the
equation (\ref{5-3}),
\begin{equation}\label{5-6}
    u^-=\frac2{\sqrt{\gamma}}\left({\arctan\sqrt{\gamma\rho^-}}-
    \arctan{\sqrt{\gamma}}\right).
\end{equation}
The ``internal'' end points $x_1^+$ and $x_2^-$ are found by
substituting the intermediate values $\rho^-$, $u^-$ into the
similarity solutions (\ref{4-5}), (\ref{4-6}),
\begin{equation}\label{5-7}
    \frac{x_1^+}{z}=u ^- - \frac{\sqrt{\rho^-}}{1+\gamma \rho^-},\qquad
    \frac{x_2^-}{z}=u ^- + \frac{\sqrt{\rho^-}}{1+\gamma \rho^-}.
\end{equation}
These points correspond to the weak discontinuities which propagate
with sound velocities $c_s(\rho^-)$ in opposite directions in the
reference frame associated with the uniform flow $u^-$.
\begin{figure}[bt]
\includegraphics[width=8cm,height=6cm,clip]{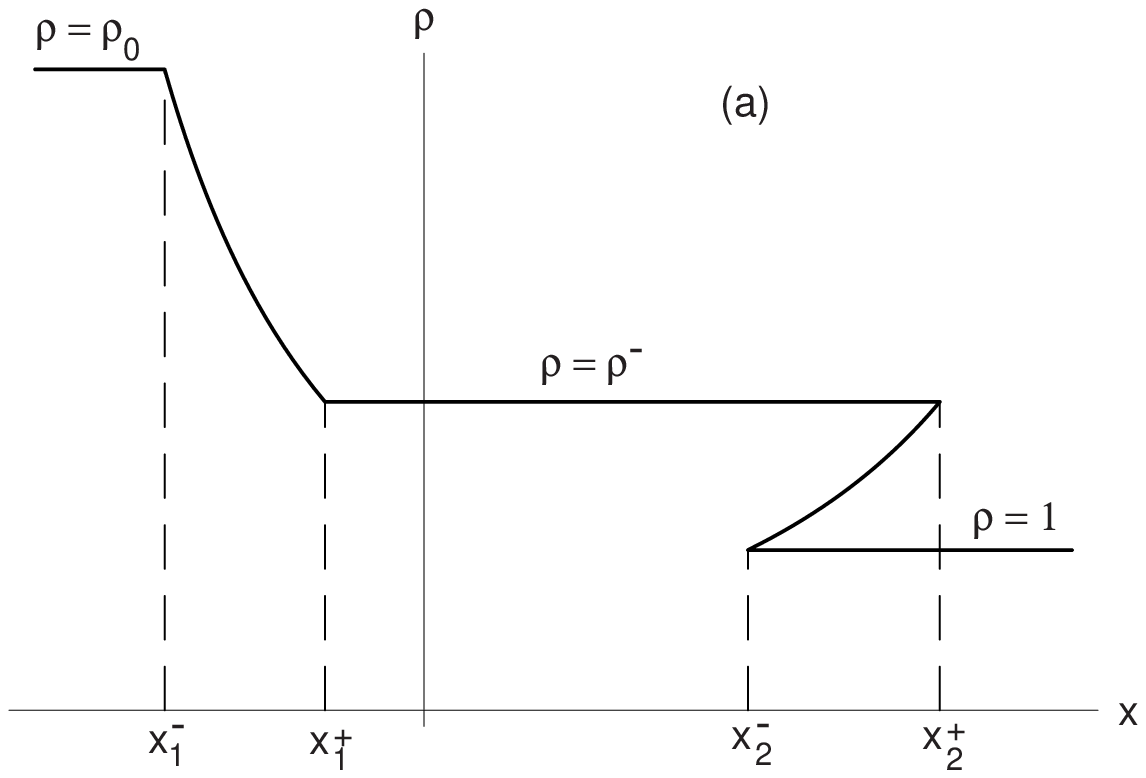}
\hspace{1cm}
\includegraphics[width=8cm,height=6cm,clip]{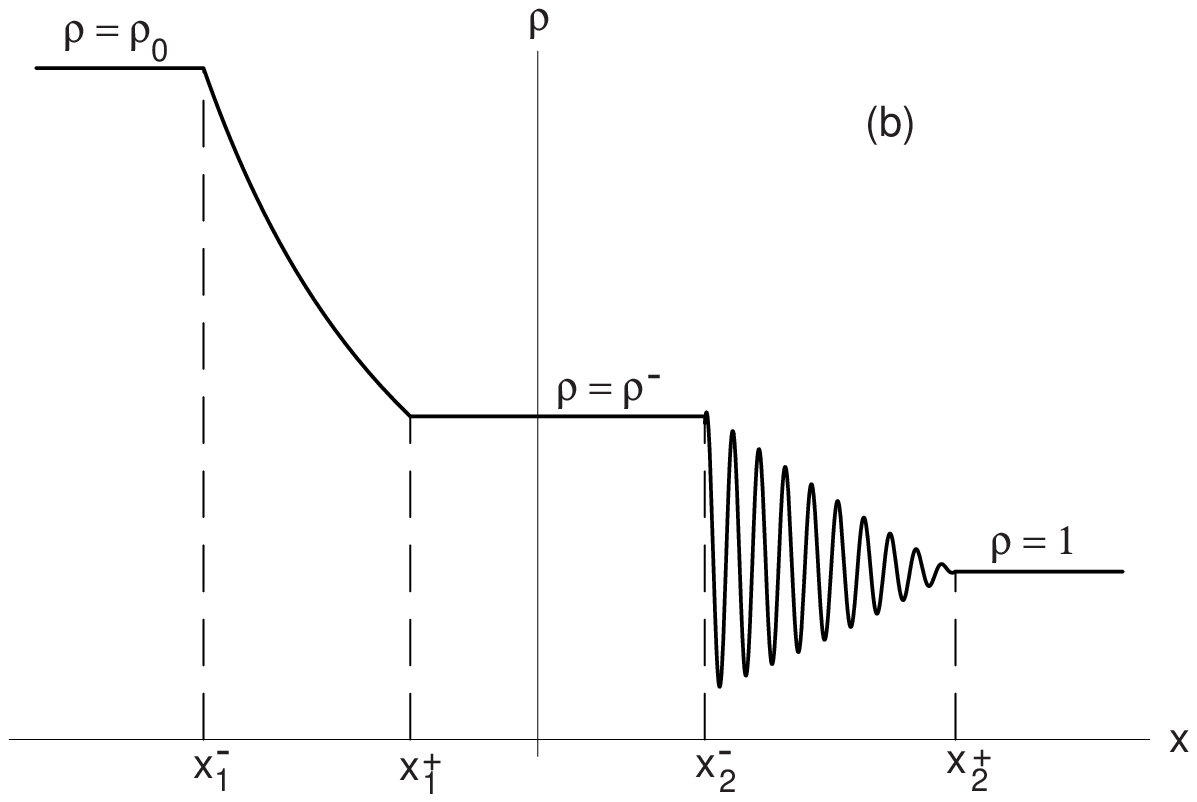}
\caption{Decay of the initial discontinuity of light intensity in a
beam propagating through a photorefractive crystal. (a)
Dispersionless approximation with non-physical region of
multi-valued intensity. (b) Schematic picture of formation of
dispersive shock due to interplay of dispersive and nonlinear
effects. The values of $x_1^{-}$ and $x_1^{+}$  are the same for (a)
and (b) while the values of $x_2^{-}$ and $x_2^{+}$ are different.}
\label{fig1}
\end{figure}
The whole structure of intensity distribution is shown in Fig.~1a.
It has the region $x_2^-<x<x_2^+$ with the three-valued intensity,
corresponding to the formal solution (\ref{4-5}), which is obviously
non-physical and its appearance serves as an indication that an
oscillating dispersive shock wave is generated in the region of
transition from $\rho=\rho^-,\,u=u^-$ to $\rho^+=1,\,u^+=0$. The
arising physical structure is shown schematically in Fig.~1b.
Importantly, the boundaries $x_{1,2}$ of the oscillatory zone by no
means coincide with those in the formal three-valued dispersionless
solution. It is remarkable, however, that in spite of such radical
qualitative and quantitative change of the flow, the values of
$\rho^-$ and $u^-$ themselves turn out to be still determined by the
previous equations (\ref{5-5}) and (\ref{5-6}). This is a
consequence of the dispersive shock jump condition which requires
that the values of the Riemann invariant
$r_-=u-(2/\sqrt{\ga})\arctan\sqrt{\ga\rho}$ at both end points of
the dispersive shock wave must be equal to each other:
\begin{equation}\label{6-1}
    \left.r_-\right|_{x_2^-}=\left.r_-\right|_{x_2^+} \, ,
\end{equation}
which gives at once Eq.~(\ref{5-6}). Since the rarefaction wave,
even in the presence of dispersion, is still described with good
accuracy by the dispersionless approximation (see \cite{gp74},
\cite{gm84} for the general linear asymptotic analysis of the
dispersive resolution of the weak discontinuities at the edges of
the rarefaction wave), we deduce that Eq.~(\ref{5-5}) obtained in
the framework of the dispersionless fluid dynamics also remains
valid. One should emphasize that, although all obtained
relationships, strictly speaking, hold only asymptotically for
sufficiently large ``times" $z$, as  we shall see from the direct
numerical solution, they hold  with good accuracy for rather
moderate $z$. The dispersive jump condition of the type (\ref{6-1})
was proposed for the first time in \cite{gm84} where it was based on
intuitive physical reasoning and the results of numerical
simulations of collisionless plasma flows. A consistent mathematical
derivation of this condition along with some important restrictions
to its applicability was given in the framework of the Whitham
theory in \cite{el,ekt}.

As was mentioned, the end points of the oscillatory region of the
dispersive shock in Fig.~1b do not coincide with the end points of
the three-valued region in Fig.~1a. Indeed, this oscillatory zone
arises due to interplay of dispersion and nonlinear effects and has
a structure  similar to that observed in the much studied integrable
defocusing NLS equation case (see \cite{gk87} -\cite{kgk04}).
Namely, near the leading edge $x_2^{+}$ the wave transforms into a
vanishing amplitude linear wavepacket and at the trailing edge
$x_2^-$ it converts into a dark soliton. Hence, the end point of the
oscillatory zone  $x_2^+$ must move with the group velocity of
linear waves $c_g=\partial \omega_0/\partial k$ calculated for some
non-zero value of $k=k^+$ in contrast to the dispersionless
approximation corresponding to $k\to 0$ (in addition to vanishing
amplitude of oscillations $a \to 0$). The end point $x_2^-$ moves
with the corresponding soliton velocity which also has nothing to do
with the dispersionless limit (note, that in the soliton limit $k
\to 0$ but the amplitude $a=a^-$ remains finite). Thus, our task is
to determine the main quantitative characteristics of the
oscillatory region of the dispersive shock---the velocities of its
end points as well as the amplitude $a^-$ of the trailing soliton at
$x=x_2^-$ and the wave number $k^+$ at the leading edge point
$x=x_2^+$.

One can observe that the oscillatory structure of the dispersive
shock wave is characterized by two different spatial scales:
the intensity oscillates very fast inside the shock but the
parameters of the fast oscillations change little in one
wavelength in $x$-direction and in one period along the beam
$z$-axis. This suggests that the oscillatory dispersive shock can be
represented as a slowly modulated nonlinear periodic wave and,
hence, we can apply the Whitham modulation theory \cite{whitham-1}
to its description. In the Whitham approach, the original equation
containing higher order $x$-derivatives is  averaged over the family
of nonlinear periodic traveling wave solutions. As a result, one
obtains a system of the first order nonlinear partial differential
equations of hydrodynamic type (i.e. linear with respect to first
derivatives) governing the slow evolution of modulations. The
modulation system does not contain any parameters of the length
dimension, so it allows one to introduce the edges $x^{\pm}_2(z)$ of
the dispersive shock wave in a mathematically consistent way, as
characteristics where matching of the ``internal'' (modulation) and
``external'' (dispersionless fluid dynamics) solutions occurs. Of
course, strictly speaking the averaged description  is valid only
when the ratio of the typical wavelength to the width of the
oscillatory zone is small. For our case of the decay of an initial
discontinuity this corresponds to a ``long-time'' asymptotic
behaviour, $z \gg 1$. However, as we shall see from the comparison
with direct numerical solution, the results of the modulation
approach turn out to be valid even for rather moderate values of
$z$.

 The modulation approach to the description of dispersive shock waves was
 realized for the first time by
Gurevich and Pitaevskii \cite{gp74} in the framework of the Korteweg
-- de Vries equation.  To put this approach into practice for the
light beam deformations in a photorefractive medium, we first have
to study periodic solutions of the equations (\ref{dh}).

\subsection{Periodic waves and solitons in photorefractive crystals}

The traveling wave solution of the system (\ref{dh}) is obtained by
the substitution $\rho=\rho(\theta)$, $u=u(\theta)$, where $\theta =
x-cz$ is the phase and $c=\hbox{constant}$ is the phase velocity. As
a result, we obtain by integrating the first equation (\ref{dh})
\begin{equation}\label{u}
u=c+\frac{A}{\rho} ,
\end{equation}
where $A$ is an arbitrary constant. Substituting (\ref{u}) into the
second equation (\ref{dh}) and performing one integration with
respect to $\theta$ we obtain an ordinary differential equation of
the second order,
\begin{equation}\label{6-2}
\frac{1}{8}\left(\frac{d\rho}{d\theta}\right)^2 = \frac{1}{4}\frac{d^2\rho}{d\theta^2}
\rho - \rho^2 f(\rho) - B\rho^2 - \frac{A^2}{2} ,
\end{equation}
where $B$ is another constant of integration. We shall seek its integral
in the form
\begin{equation}\label{sub}
\left(\frac{d\rho}{d\theta}\right)^2 = a_1 \rho \int f(\rho)d\rho +
a_2\rho^2+a_3\rho+a_4 ,
\end{equation}
where $a_1$, $a_2$, $a_3$, and $a_4$ are the constant coefficients
to be found. Substituting (\ref{sub}) into (\ref{6-2}) we find, with
the account of the specific dependence $f(\rho)$, the eventual form
of the sought integral,
\begin{equation}\label{trav}
\left(\frac{d\rho}{d\theta}\right)^2=-\frac{8\rho}{\gamma^2}\ln
(1+\gamma \rho)+ \left(a_2+\frac{8}{\gamma}\right)\rho^2 + a_3\rho
+a_4 \equiv Q(\rho) .
\end{equation}
Here $a_2$, $a_3$ and $a_4$  are arbitrary constants two of which are
connected with $A$ and $B$ by the relations
\begin{equation}\label{6-3}
     a_2=8B, \quad a_4=-4A^2,
\end{equation}
and $a_3$ is an additional constant so that (\ref{trav}) is indeed
the first integral of Eq.~(\ref{6-2}).
 We denote the roots of the equation $Q(\rho)=0$
as $e_1 \le e_2 \le e_3$. Then the density oscillations in the
traveling wave occur between $e_1$ and $e_2$. The amplitude of the
wave is then given by $a=e_2-e_1$. The small-amplitude linear wave
configuration corresponds to $e_1 \to e_2$ while for solitons we
have $e_2=e_3$. By imposing the periodicity condition
$\rho(\theta)=\rho (\theta + 2\pi/k)$ we find the wave number $k$ of
the traveling wave in the form of the integral
\begin{equation}\label{k}
k= \pi \left (\int_{e_1}^{e_2}
\frac{d\rho}{\sqrt{Q(\rho)}}\right)^{-1}.
\end{equation}

While the equation (\ref{trav})
cannot be integrated in closed form, it is not difficult to find the
relationships characterizing its special solution in the form of a
dark soliton. For this solution we must have the following boundary
conditions satisfied at infinity:
\begin{equation}\label{sbc}
\rho \to \rho_b, \quad u \to u_b , \quad
d\rho/d\theta \to 0 , \quad d^2\rho/d\theta^2 \to 0
\qquad \text{for}\quad |\theta| \to \infty,
\end{equation}
plus the condition $d\rho/d\theta=0$ at $\rho=\rho_m  \le \rho_b$, where
$\rho_m$ is the value of the ``density" in the minimum of the dark
soliton and $\rho_b$ is the ``background'' intensity.
Applying these conditions to (\ref{u}), (\ref{trav}) we
obtain, after simple algebra, the expressions for the coefficients
in (\ref{trav}) for the soliton configuration,
\begin{equation}\label{ACD}
\begin{split}
a_2&=-\frac{8\rho_b}{1+\ga\rho_b}-4(u_b - c)^2, \\
a_3&=\frac{8}{\gamma^2}\ln(1+\rho_b) -
\frac{8\rho_b}{\gamma(1+\gamma\rho_b)} +
\frac{4(u_b-c)^2(\rho_m^2+\rho_b^2)}{\rho_b},\\
a_4&=-4(u_b-c)^2\rho_b^2.
\end{split}
\end{equation}
The curves $Q(\rho)$ in a ``soliton configuration'' for several
values of $\ga$ are shown in Fig.~2.
\begin{figure}[bt]
\includegraphics[width=8cm,height=6cm,clip]{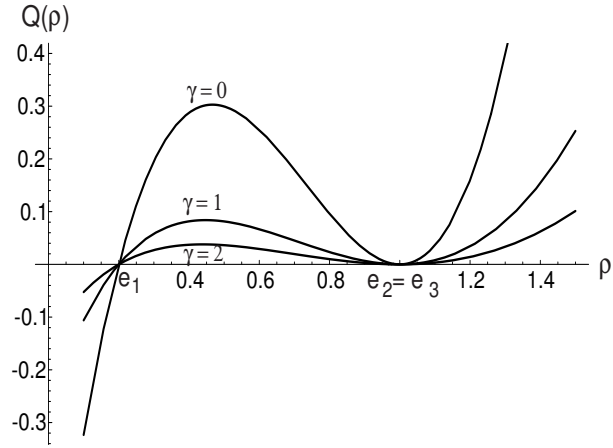}
\caption{Plots of the function $Q(\rho)$ corresponding to
$\rho_b=1$, $\rho_m=0.2$ and different values of $\ga$ and
$\rho_b=1,$ $\rho_m=0.2,$ so that $e_1=0.2,$ $e_2=e_3=1$. }
\label{fig2}
\end{figure}
The condition that in the soliton limit $\rho_b$ is a double zero
of the function $Q(\rho)$, that is $dQ(\rho)/d\rho=0$ at $\rho=\rho_b$,
yields the relationship between the soliton velocity $c$ and the amplitude
$a=\rho_b-\rho_m$ for given $\rho_b$, $u_b$:
\begin{equation}\label{speedamp}
(c-u_b)^2=\frac{2\rho_m}{\gamma a}\left[\frac{1}{\gamma a}\ln
\frac{1+\gamma\rho_b}{1+\gamma \rho_m}- \frac{1}{1+\gamma \rho_b}
\right].
\end{equation}
The dependence of the soliton velocity on the saturation parameter $\ga$
is shown in Fig.~3.
\begin{figure}[bt]
\includegraphics[width=8cm,height=6cm,clip]{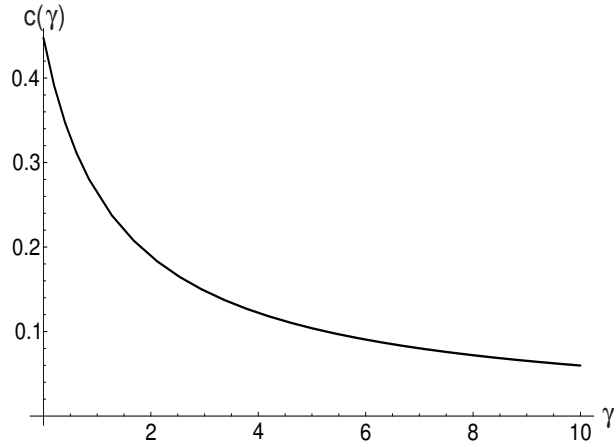}
\caption{The plot of soliton velocity as a function of the
saturation parameter $\ga$. The other parameters are: $\rho_b=1$,
$\rho_m=0.2$. } \label{fig3}
\end{figure}

For future analysis it is important to introduce one more
parameter---the inverse half-width $\kappa$ of the soliton--- using
the exponential decay of the intensity $\rho_b-\rho$ as
$|\theta|\to\infty$:
\begin{equation}\label{7-1}
    \rho_b-\rho\propto \exp(-\kappa|\theta|),\quad |\theta|\to \infty.
\end{equation}
To find the relationship between $\kappa$ and other parameters we
take the series expansion of $Q(\rho)$ for small values of
$\rho'=\rho_b-\rho$ and find
$(d\rho'/d\theta)^2=(1/2)(d^2Q/d\rho^2)_{\rho_b}(\rho')^2=
\kappa^2(\rho')^2$; hence
\begin{equation}\label{7-2}
    \kappa=\left(\frac12\left.\frac{d^2Q}{d\rho^2}\right|_{\rho_b}\right)^{1/2}=
    \left[\frac{8\rho_m+4\ga \rho_b(\rho_b+\rho_m)}{\ga(\rho_b-\rho_m)(1+\ga\rho_m)^2}-
    \frac{8\rho_m}{\ga^2(\rho_b-\rho_m)^2}\ln\frac{1+\ga\rho_b}{1+\ga\rho_m}
    \right]^{1/2}.
\end{equation}
The dependence of $\kappa$ on $\ga$ is shown in Fig.~4.
\begin{figure}[bt]
\includegraphics[width=8cm,height=6cm,clip]{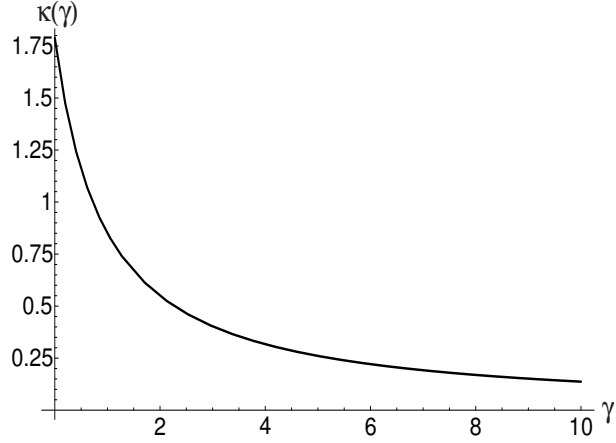}
\caption{The plot of inverse half-width $\kappa$ of photorefractive
soliton as a function of saturation parameter $\ga$. The other
parameters are: $\rho_b=1$, $\rho_m=0.2$. } \label{fig4}
\end{figure}

The profile of the intensity $\rho(\theta)$ is determined by
the integral (see Eq.~(\ref{trav}))
\begin{equation}\label{7-3}
    \theta=\int_{\rho_m}^{\rho}\frac{d\rho}{\sqrt{Q(\rho)}},
\end{equation}
where it is assumed that the intensity $\rho$ takes the minimal
value $\rho=\rho_m$ at $\theta=0$ which determines the
integration constant. The wave form of a dark soliton for different values of
the parameter $\gamma$ is shown in Fig.~5.
\begin{figure}[bt]
\includegraphics[width=8cm,height=6cm,clip]{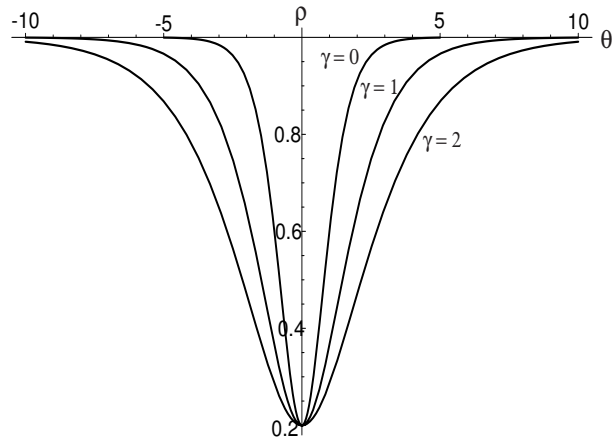}
\caption{Profiles of the intensity in photorefractive solitons for
values of $\ga=0,\,1,\,2$. The other parameters are: $\rho_b=1$,
$\rho_m=0.2$. } \label{fig5}
\end{figure}

For $\gamma \ll 1$ we have asymptotic expansions (for simplicity
we take $u_b=0$)
\begin{equation}\label{as}
c=\sqrt{\rho_m}\left(1 - \frac{\ga}3
(2\rho_b+\rho_m)\right)+\mathcal{O}(\gamma^2) .
\end{equation}
\begin{equation}\label{8-1}
    \kappa=2\sqrt{\rho_b-\rho_m}\left[1-\frac{\ga}3(3\rho_b+\rho_m)\right]+\mathcal{O}(\gamma^2),
\end{equation}
and for $\ga\gg1$ other expansions
\begin{equation}\label{8-2}
    c=\frac{\sqrt{2[\rho_m(\ln(\rho_b/\rho_m)-1)+\rho_m^2/\rho_b]}}{(\rho_b-\rho_m)\ga}
    +\mathcal{O}(\gamma^{-2}),
\end{equation}
\begin{equation}\label{8-3}
    \kappa=\frac{2\sqrt{\rho_b-2\rho_m\ln(\rho_b/\rho_m)-\rho_m^2/\rho_b}}{(\rho_b-\rho_m)\ga}
    +\mathcal{O}(\gamma^{-3}) .
\end{equation}

One can see that the leading terms in (\ref{as}) and (\ref{8-1})
agree with the well-known dependencies for dark solitons of the NLS
equation \cite{gk87}.

The particular case of soliton solution with $\rho_m=0$, $u_b=0$ (hence $c=0$)
in photorefractive media has been found in \cite{hou}.

\subsection{Dispersive shock wave}

The general periodic solution of the photorefractive equation
depends on the fast phase variable $\theta$ and is characterized
by four parameters $e_1,\, e_2,\, e_3, c$, where $e_j, \ j=1,2,3$
are the zeroes of the function $Q(\rho)$ (\ref{trav}), which
determine the profile of the intensity, and $c$ is the phase
velocity. In a modulated wave, these four parameters become slow
variables of $x$ and $z$. In the Whitham theory \cite{whitham-1}, it
is postulated that this slow evolution (modulation) $e_j(x,z)$,
$c(x,z)$ can be found from the conservation laws of the dispersive
equation averaged over fast oscillations with respect to the phase
variable $\theta$. An additional modulation equation naturally
arises as the wave number conservation law $k_z+\omega_x=0$ and
essentially represents a condition of the existence of a slowly
modulated periodic wave (see, for instance, \cite{whitham-1}).
Several averaging procedures have been proposed yielding equivalent
results for various physical systems (see \cite{ri}) so the Whitham
modulation theory can be now considered as quite well established.
As a result, using the original procedure of averaging conservation
laws, the Whitham system for the GNLS equation can be obtained in
the following general form
\begin{equation}\label{wh}
(\overline {P_i}(e_1, e_2, e_3,c))_z + (\overline{Q_i}(e_1, e_2,
e_3,c))_x =0\, , \qquad i=1,2,3 ,
\end{equation}
\begin{equation}\label{cw}
(k(e_1, e_2, e_3, c))_z + (\omega(e_1, e_2, e_3,c))_x=0 \, , \qquad
\omega=kc.
\end{equation}
Here $P_1=\rho$, $P_2=u$, $P_3=\rho u$ are the conserved
``densities'' of the GNLS equation (\ref{1-8}) and $Q_i, \ i=1,2,3,$
are the corresponding ``fluxes''. The averaging is performed over
the periodic family (\ref{u}), (\ref{trav}) according to
\begin{equation}\label{aver}
\overline f (e_1,e_2,e_3, c) = \frac{k}{\pi} \int
\limits_{e_1}^{e_2} \frac{f(\rho; e_1,e_2,e_3, c)}{\sqrt{Q(\rho)}} d
\rho \, .
\end{equation}
Now the system (\ref{wh}), (\ref{cw})  is, in principle, completely
defined.

The modulation system (\ref{wh}), (\ref{cw}) being the system of
hydrodynamic type  can be hyperbolic (real characteristic velocities
-- modulationally stable case) or elliptic (complex characteristic
velocities -- modulationally unstable case). It is known very well
(see \cite{forest}, \cite{pavlov}, \cite{kamch2000}) that for the
defocusing NLS equation, which is an integrable particular case of
the the GNLS equation (\ref{1-5}), the modulation system is strictly
hyperbolic. Our numerical simulations show that traveling waves in
the GNLS equation are modulationally stable and this suggests that
the corresponding Whitham system is hyperbolic as well. So, in what
follows, we shall assume hyperbolicity of the Whitham system, which
will allow us  to use some arguments of classical characteristics
theory \cite{lanlif, courant, whitham-1}.

Now, to describe analytically the dispersive shock wave as a whole,
we have to solve four modulation equations (\ref{wh}), (\ref{cw})
for the slowly varying parameters $e_1, e_2, e_3$ and $c$ of the
periodic solution. These equations  must be equipped with special
matching conditions to guarantee continuity of the mean flow at the
free boundaries $x^{\pm}(z)$ defining the edges of the dispersive
shock wave. In view of the numerically established qualitative
spatial structure of the photorefractive dispersive shock wave (see
Fig.~1b) we require that
\begin{equation}\label{lead}
\hbox{at} \quad x=x^{+}(z): \qquad a=0\, , \quad \overline \rho=
\rho^+\, , \quad \overline u = u^+ \, ,
\end{equation}
\begin{equation}\label{trail}
\hbox{at} \quad x=x^{-}(z): \qquad k=0\, , \quad \overline \rho=
\rho^-\, , \quad \overline u = u^- \, ,
\end{equation}
where $ x^+\equiv x_2^+$ (from now on we shall omit the subscript 2
in $x_2^-$ and $x_2^+$). The dependencies of $\overline \rho$,
$\overline u$, $k$, $a$ on $e_1, e_2, e_3, c$ are defined by
(\ref{aver}) and formulae of Section III.B, and the pairs
$(\rho^{-}$, $u^{-})$ and $(\rho^{+}$, $u^{+})$ represent the
solution of the dispersionless approximation (\ref{4-1}) evaluated
at the trailing and leading edges  of the dispersive shock wave
respectively. The edges
 $x^{\pm}(z)$ of the dispersive shock wave represent
free boundaries defined by the kinematic boundary conditions with
clear physical meaning explained in Section III.A:
\begin{equation}\label{kinematic}
\frac{d x^+}{dz}=c_g(\rho^+, u^+, k^+)\, , \qquad \frac{d
x^-}{dz}=c_{sol}(\rho^-, u^-, a^-) \, ,
\end{equation}
where $c_g(\rho^+,  u^+, k ) =\partial \omega_0/\partial k $ is the
group velocity of the linear wave packet with the dominant
wavenumber $k$ propagating against the hydrodynamic background
$\rho^+$, $u^+$ (see (\ref{dr}) for the linear dispersion relation
$\omega = \omega_0(\rho_0, u_0, k)$) and $c_{sol}(\rho^-, u^-, a)$
is the velocity of the dark soliton with amplitude $a$ propagating
against the background $\rho^-$, $u^-$ (see (\ref{speedamp}) for the
dependence of the soliton velocity on its amplitude). Of course, the
values of the wavenumber $k^+$ at the leading edge and the amplitude
$a^-$ of the trailing dark soliton are both to be determined, so the
determination of the edges $x^{\pm}(z)$ represents a part of this
nonlinear boundary value problem.

Following the pioneering work of Gurevich and Pitaevskii \cite{gp74}
on the dispersive shock wave description in the framework of the KdV
equation, the effective methods for treatment of such problems have
been developed for the whole class of evolution equations which
share with the KdV equation the unique property of complete
integrability (see, e.g., \cite{kamch2000}). On the level of the
Whitham equations, one of the manifestations of integrability is the
presence of the full system of Riemann invariants, an event
generally highly unlikely for the systems of hydrodynamic type with
number of equations exceeding two. In particular, the NLS equation
(\ref{1-6}) belongs to this class, and the corresponding theory of
dispersive shock formation was developed in the papers
\cite{gk87,eggk95, ek95, jin,tian,kku, bk} and successfully applied
to the description of shocks in nonlinear optics \cite{kodama} and
Bose-Einstein condensates \cite{kgk04,hoefer}. However, the
photorefractive equation (\ref{1-5}) is not completely integrable
and therefore the methods based on the presence of rich underlying
algebraic structure of such equations cannot be applied here.
Nevertheless, as was shown in \cite{el}-\cite{ekt}, the main
quantitative characteristics of the dispersive shock wave can be
derived using the general properties of the Whitham equations
(\ref{wh}), (\ref{cw}) reflecting their origin as certain averages,
and here we shall apply this method to the description of dispersive
shock waves in photorefractive media. To be specific, we shall be
interested in the locations of the edges of the dispersive shock
wave and in the amplitude of the largest (deepest) soliton at the
trailing edge, the parameters that are usually observed in
experiment.

The method of Refs.~\cite{el}-\cite{ekt}, which will be used below,
is formulated most conveniently in terms of the physical modulation
parameters $\overline{\rho},\,\overline{u},\,k,\,a$ appearing in the
matching conditions (\ref{trail}), (\ref{lead}). The key of the
method lies in the fact that the modulation system (\ref{wh}),
(\ref{cw}) dramatically simplifies in the cases ($a=0, \ k \ne 0$)
and ($k=0, \ a \ne 0$) corresponding to the limiting wave regimes
realized at the boundaries of the dispersive shock wave.

\subsubsection{Leading edge}

At the leading edge $x=x^+(z)$  the amplitude of oscillations
vanishes, $a=0$. Since the Whitham averaging procedure remains valid
for the case $a=0$ (averaging over the periodic family with
vanishing amplitude), then we conclude that the Whitham system must
admit an {\it exact} reduction at $a=0$ and, therefore, the system
of four Whitham equations must reduce here to only three equations.
Now, if the amplitude of oscillations vanishes, then the average of
a function of the oscillating variable equals to the same function
of the averaged variable: $\overline{F(\rho,u)}=F(\overline{\rho},
\overline{u})$. Thus, when $a=0$ the Whitham system must agree with
the dispersionless approximation (\ref{4-1}) describing large-scale
non-oscillating flows,  i.e. the modulation equations for $\bar
\rho$, $\bar u$, $a$ reduce to
\begin{equation}\label{10-1}
  a=0, \qquad   \overline{\rho}_z+(\overline{\rho}\,\overline{u})_x=0,\quad
    \overline{u}_z+\overline{u}\,\overline{u}_x+f'(\overline{\rho})\overline{\rho}_x=0.
\end{equation}
We note that this reduction of the Whitham equations is also
consistent with the  matching condition (\ref{lead})
 at the leading edge of the dispersive
shock wave  where $a=0$ and which requires that the solution of the
Whitham equations must match with the solution of the equations of
the dispersionless approximation.  Of course, equations (\ref{10-1})
can be derived directly from the modulation equations (\ref{wh}) by
passing in them to the limit $a=e_2-e_1 \to 0$ (see, for instance,
\cite{egs} for the corresponding calculation in the context of fully
nonlinear shallow-water waves), however, validity of (\ref{10-1})
appears to be obvious from the presented qualitative reasoning.

To complete the zero-amplitude reduction of the modulation system we
need to pass to the same limit as $a \to 0$ in the ``number of
waves'' conservation law (\ref{cw}) in which we assume the
aforementioned change of variables $(e_1,e_2,e_3,c) \mapsto
(\overline{\rho},\,\overline{u},\,k,\,a)$,
\begin{equation}\label{10-2}
    k_z+(\omega(\overline{\rho},\overline{u}, k,a))_x=0,\quad
    \omega=kc.
\end{equation}
As a result, we get
\begin{equation}\label{10-3}
    k_z+(\omega_0(\overline{\rho},\overline{u}, k))_x=0,
\end{equation}
where
\begin{equation}\label{10-4}
    \omega_0(\overline{\rho},\overline{u}, k))=k\left(\overline{u}+
    \sqrt{\frac{\overline{\rho}}{(1+\ga\overline{\rho})^2}+\frac{k^2}4}
    \right)
\end{equation}
is the dispersion relation (\ref{dr}) of linear waves propagating
about slowly varying background with locally constant values of
$\overline{\rho}$ and $\overline{u}$ (here we restrict ourselves
with right-propagating waves). Equations (\ref{10-1}), (\ref{10-3})
comprise a closed system which represents an exact zero-amplitude
reduction of the full Whitham system (\ref{wh}), (\ref{cw}) (see
\cite{el}, \cite{ekt} for a detailed justification of this reduction
for a class of weakly dispersive nonlinear systems) and, as we shall
see, its analysis with an account of boundary conditions
(\ref{lead}), (\ref{trail}) yields the necessary information about
the leading edge $x=x^+(z)$ of the dispersive shock wave.

Now we observe that the ``ideal'' hydrodynamic equations
(\ref{10-1}) are decoupled from (\ref{10-3}) and thus, can be solved
independently for $\overline \rho (x,z)$, $\overline u (x,z)$.
However, since the values of $\overline \rho$ and $\overline u$ at
$a=0$ are subject to boundary conditions (\ref{lead}), one should
take into account the restriction on the admissible values of
$\overline \rho$ and $\overline u$ at the boundaries of dispersive shock
wave  imposed by the simple-wave transition condition (\ref{6-1}).
Since this restriction is consistent with the equations
(\ref{10-1}), it can be incorporated directly into the reduced
modulation system by putting
\begin{equation}\label{simp}
\overline u = \frac2{\sqrt{\gamma}}\left({\arctan\sqrt{\gamma
\overline \rho}}-
    \arctan{\sqrt{\gamma}}\right).
\end{equation}
Substitution of (\ref{simp}) into system (\ref{10-1}), (\ref{10-3})
further reduces it to only two differential equations
\begin{equation}\label{whr}
 \qquad \overline \rho _z + V_+(\overline \rho)\overline \rho
_x =0\, , \qquad k_z+(\Omega(\overline \rho, k))_x =0 \, ,
\end{equation}
where
\begin{equation}\label{}
V_+(\overline \rho) =  \frac{2}{\sqrt{\gamma}}(\arctan{\sqrt{\gamma
\overline \rho}}- \arctan{\sqrt{\gamma}}) + \frac{\sqrt{\overline
\rho}}{1+\gamma \overline \rho}\,\, ,
\end{equation}
\begin{equation}\label{12-1}
    \Omega(\overline \rho,k)=\omega_0(\overline \rho,
    \overline u(\overline \rho),k)=k\left[\frac2{\sqrt{\ga}}
    \left(\arctan\sqrt{\ga \overline \rho}-
    \arctan\sqrt{\ga}\right)+\sqrt{\frac{\overline \rho}{(1+\ga \overline \rho)^2}+
    \frac{k^2}4}\right].
\end{equation}

The system (\ref{whr}) has two families of characteristics:
\begin{equation}\label{ch1}
\frac{dx}{dz}=V_+(\overline \rho)
\end{equation}
and
\begin{equation} \label{ch2}
 \frac{dx}{dz}=\dfrac{\partial
\Omega (\overline \rho, k)}{\partial k}\, .
\end{equation}
The family (\ref{ch1}) is completely determined by the simple-wave
evolution of the function $\overline \rho (x,z)$ according to the
dispersionless approximation of the GNLS equation. This family
transfers ``external'' hydrodynamic data into the dispersive shock
wave region and does not depend on the oscillatory structure.
Contrastingly, the behaviour of the characteristics belonging to the
family (\ref{ch2}) depends on both $\overline \rho $ and $k$.
Comparison of the definition (\ref{kinematic}) of the leading edge
$x^+(z)$ with (\ref{ch2}) with the account of (\ref{12-1}) shows
that the leading edge of the dispersive shock wave represents a
characteristic belonging to the family (\ref{ch2}). Now, since the
system (\ref{whr}) consists of two equations, then according to
general properties of characteristics of nonlinear hyperbolic
systems of partial differential equations (see, for instance,
\cite{courant}, \cite{whitham-1}, \cite{lanlif}), one cannot specify
two values $k$ and $\overline \rho$ independently on one
characteristic, so the admissible combinations of $\overline \rho$
and $k$ at the leading edge of the dispersive shock wave are
determined by a characteristic integral
 of the reduced modulation system (\ref{whr}).

To this end, we substitute $k=k(\overline \rho)$ into (\ref{whr}) to
obtain at once
\begin{equation}\label{12-2}
 a=0: \qquad    \frac{dk}{d \overline \rho}=\frac{\prt\Omega/\prt \overline \rho}{V_+-\prt\Omega/\prt
    k} \qquad \hbox{on} \qquad \frac{dx}{dz}=\frac{\partial \Omega}{\partial
    k}\, .
\end{equation}
The above ordinary differential equation for $k$ must be solved with
the initial condition $k(\rho^-)=0$. Indeed, since the equation
(\ref{12-2}) was derived for the case $a=0$ it must remain valid in
the case of the dispersive shock wave of zero intensity, so the
dependence $k(\overline \rho)$ should correctly reproduce the zero
wavenumber condition at the trailing edge where $\overline \rho =
\rho^-$ (see (\ref{trail})).

By introducing the variable
\begin{equation}\label{12-3}
    \alpha=\sqrt{1+\frac{k^2(1+\ga \overline \rho)^2}{4 \overline
    \rho}}\, ,
\end{equation}
instead of $k$, in (\ref{12-2}), and using Eq.~(\ref{12-1}),  we
arrive at the ordinary differential  equation
\begin{equation}\label{12-4}
    \frac{d\alpha}{d \overline \rho}=-\frac{(1+\alpha)[1+3\ga \overline \rho+2\alpha(1-\ga \overline \rho)]}
    {2 \overline \rho(1+\ga \overline \rho)(1+2\alpha)}.
\end{equation}
with the initial condition
\begin{equation}\label{12-5}
    \alpha(\rho^-)=1 \, ,
\end{equation}
where $\rho^-$ is determined in terms of the initial discontinuity
(\ref{step}) by Eq.~(\ref{5-5}). Once the solution $\alpha
(\overline \rho)$ is found, the wave number $k^+$ at the leading
edge, where $\overline \rho = \rho^+=1$, is determined from
(\ref{12-3}) as
\begin{equation}\label{12-6}
    k^+=k (1) = \frac{2\sqrt{\alpha^2(1)-1}}{1+\ga}\, .
\end{equation}
The velocity of propagation of the leading edge is defined by the
kinematic condition (\ref{kinematic}), which , with an account of
(\ref{ch2}), assumes the form
\begin{equation}\label{12-7}
s^+=\frac{dx^+}{dz}=\frac{\partial \Omega}{\partial k}
(1,k^+)=\frac1{1+\ga}\left(2\alpha(1)-\frac1{\alpha(1)}\right).
\end{equation}

For the NLS equation case, i.e. when $\ga=0$, the expression for
$s^+$ in terms of the density jump across the dispersive shock wave
can be obtained explicitly: the equation
\begin{equation}\label{13-1}
    \frac{d\alpha}{d \overline \rho}=-\frac{1+\alpha}{2 \overline \rho},\qquad \alpha(\rho^-)=1
\end{equation}
is readily integrated to give
\begin{equation}\label{13-2}
    \alpha(\overline \rho)=2\sqrt{\frac{\rho^-}{\overline \rho}}-1
\end{equation}
and thus
\begin{equation}\label{13-3}
    s^+=\frac{8\rho^--8\sqrt{\rho^-}+1}{2\sqrt{\rho^-}-1}\quad
    \text{for}\quad \ga=0
\end{equation}
in agreement with known results \cite{gk87}.

For small values of the saturation parameter $\gamma\ll1$ one can
find the correction to this formula with the use of
Eqs.~(\ref{12-4}) and (\ref{12-7}). Indeed, if we introduce
$\alpha=\alpha_0+\alpha_1$, where $\alpha_0$ is given by
Eq.~(\ref{13-2}) and $\alpha_1$ has the order of magnitude of
$\gamma$, then the series expansion of Eq.~(\ref{12-4}) yields the
differential equation for the correction $\alpha_1$:
\begin{equation}\label{10a-1}
    \frac{d\alpha_1}{d \overline \rho}=-\frac{\alpha_1}{2\overline \rho}+\frac{8\sqrt{\rho^-/ \overline \rho}-6}
    {4\sqrt{\rho^-/\overline \rho}-1}\sqrt{\frac{\rho^-}{\overline
    \rho}}\,\gamma \, ,
\end{equation}
which can be easily solved with account of the initial condition
$\alpha_1(\rho^-)=0$ to give
\begin{equation}\label{10a-2}
    \alpha_1(1)=2\gamma\sqrt{\rho^-}\left\{1-\rho^-+
    64\left[\ln\frac{4\sqrt{\rho^-}-1}{3\sqrt{\rho^-}}+\frac{1-\sqrt{\rho^-}}
    {4\sqrt{\rho^-}}+\frac{1-\rho^-}{32\rho^-}\right]\right\}.
\end{equation}
Then substitution of this expression into Eq.~(\ref{12-7}) gives an
explicit approximate formula for $s^+$:
\begin{equation}\label{10a-3}
    s^+=\frac{8\rho^--8\sqrt{\rho^-}-1}{2\sqrt{\rho^-}-1}(1-\gamma)
    +\left[2-\frac1{(2\sqrt{\rho^-}-1)^2}\right]\alpha_1(1),
\end{equation}
which is correct for small $\gamma$ as long as $\alpha_1(1)\ll1$.

\subsubsection{Trailing edge}

In the vicinity of the trailing edge $x=x^-(z)$ the photorefractive
dispersive shock wave represents a sequence of weakly interacting
dark solitons propagating on the slowly varying background
$\overline{\rho},\,\overline{u}$. Since  one has  $k \to 0$ as $x
\to x^-$, we shall be interested in passing to a soliton limit in
the modulation system (\ref{wh}), (\ref{cw}). Instead of performing
this limiting passage by a direct calculation (which can be quite
involved technically), we shall invoke the reasoning similar to that
used in the study of the zero-amplitude regime to investigate a
reduced modulation system as $k \to 0$.

In limit as $k \to 0$, the distance between solitons (i.e., a
wavelength $2\pi/k$) tends to infinity, so the contribution of
solitons into the averaged flow $\overline \rho$, $\overline u$
vanishes, and, similarly to the case of the vanishing amplitude, we
have $\overline{F(\rho,u)}=F(\overline{\rho}, \overline{u})$. Hence,
we arrive again at the ideal hydrodynamics system (\ref{10-1}) for
$\overline \rho$, $\overline u$. Next, using the arguments identical
to those used earlier for the case $a=0$ but applied now to the case
$k=0$ we conclude that, for the matching condition (\ref{trail}) at
the trailing edge to be consistent with the simple-wave transition
condition (\ref{6-1}) we should incorporate the relation
(\ref{simp}) into the reduced as $k \to 0$ modulation system to
obtain the same equation for $\overline \rho$ (see (\ref{whr})),
which we reproduce one more time:
\begin{equation}\label{sw2}
\overline \rho _z + V_+(\overline \rho)\overline \rho _x =0\, .
\end{equation}

Now we need to pass to the limit as $k \to 0$ the wave conservation
law.  This  limiting transition, unlike that as $a \to 0$, is a
singular one, so it requires a more careful consideration. First we
note that the wave conservation law  is identically satisfied for
$k=0$ so we need to take into account higher order terms in the
expansion of (\ref{10-2}) for small $k$. Following \cite{el},
\cite{ekt} we introduce a ``conjugate wave number''
\begin{equation}\label{15-2}
    \widetilde{k}=\pi\left(\int_{e_2}^{e_3}
    \frac{d\rho}{\sqrt{-Q(\rho)}}\right)^{-1}
\end{equation}
instead of the amplitude $a$ and the ratio $\Lambda=k/\widetilde{k}$
instead of the original wave number $k$, so that the parameters
$(\overline{\rho},\,\overline{u},\, \Lambda, \, \widetilde{k})$
provide a new set of the modulation parameters which is convenient
for consideration of the vicinity of the soliton edge of a
dispersive shock. The variable $\widetilde{k}$ can be considered as
a wave number of oscillations of the variable $\rho$ in the interval
$e_2 \le \rho \le e_3$ governed by the ``conjugate'' traveling wave
equation
\begin{equation}\label{15-3}
    \left(\frac{d\rho}{d\widetilde{\theta}}\right)^2=-Q(\rho)\, ,
\end{equation}
where $Q(\rho)$ is defined in Eq.~(\ref{trav}) and $\widetilde
\theta$ is a new phase variable. In the soliton limit $e_2\to e_3$
we can expand $Q(\rho)$ in the vicinity of its minimum point
$\overline{\rho}=e_2=e_3$ so that Eq.~(\ref{15-3}) takes the form of
the ``energy conservation law'' of the harmonic oscillator,
$$
\frac12\left(\frac{d\rho}{d\widetilde{\theta}}\right)^2+
\frac14\left.\frac{d^2Q}{d\rho^2}\right|_{\overline \rho}
(\rho-\overline{\rho})^2=Q(\overline \rho ).
$$
Then comparison with Eq.~(\ref{7-2}) shows that in this limit
\begin{equation}\label{15-5}
 \widetilde{k}=\sqrt{\frac12\left.\frac{d^2Q}{d\rho^2}\right|_{\overline \rho}}
    =\kappa \, ,
\end{equation}
which explains the physical meaning of the variable $\widetilde{k}$
in the limit  we are interested in. This analogy can be amplified by
noticing that Eq.~(\ref{15-3}) can be viewed as the traveling wave
equation  corresponding to the ``conjugate'' GNLS equation obtained
from (\ref{1-5}) by replacing the variables  $x$ and $z$ by $ix$ and
$iz$ respectively so that $\theta$ in (\ref{trav}) is replaced by $i
\widetilde \theta$ what leads to the change of sign in (\ref{trav})
transforming this equation to Eq.~(\ref{15-3}). Now, the same
transformation maps a harmonic wave $\exp[i(kx-\omega z)]$ to the
tails of the soliton solution $\exp[\pm\kappa(x-c_{sol}z)]$, that is
in the soliton limit the conjugate frequency $\widetilde{\omega}_0$
can be obtained from the harmonic dispersion relation by a
substitution
\begin{equation}\label{15-7}
    i\widetilde{\omega}_0=\omega_0(i\kappa).
\end{equation}
Actually, this fact is well known and can be used for calculation of
the dependence of the soliton velocity $c_{sol}=\widetilde \omega_0
/\kappa$ on its inverse half-width $\kappa$ from the dispersion
relation for linear waves (see, e.g., \cite{dkn}). Thus, for
photorefractive dark solitons propagating along the slowly varying
background $\overline \rho$, $\overline u$ we have the conjugate
dispersion relation
\begin{equation}\label{11-2}
    \widetilde{\omega}_0(\overline{\rho},\overline{u}, \kappa))=\kappa
    \left(\overline{u}+
    \sqrt{\frac{\overline{\rho}}{(1+\ga\overline{\rho})^2}-\frac{\kappa^2}4}
    \right) \, ,
\end{equation}
which, after substitution of the simple-wave relation (\ref{simp}),
assumes the form (cf. (\ref{12-1}))
\begin{equation}\label{13-4}
    \widetilde{\Omega}(\overline \rho,\kappa)=
    \widetilde \omega_0 (\overline \rho, \overline u (\overline \rho),
    \kappa)=\kappa\left[\frac2{\sqrt{\ga}}
    \left(\arctan\sqrt{\ga \overline \rho}-
    \arctan\sqrt{\ga}\right)+\sqrt{\frac{ \overline \rho}{(1+\ga \overline \rho)^2}-
    \frac{\kappa^2}4}\right] \, .
\end{equation}
Now we are ready to study the asymptotic expansion as $k \to 0$ of
the wave conservation law (\ref{10-2}). First we substitute
$k=\Lambda \widetilde{k}$ into Eq.~(\ref{10-2}) to obtain
\begin{equation}\label{cw2}
\tilde k  \Lambda_z + \tilde \omega \Lambda_x+ \Lambda ( \tilde k_z
+ \tilde \omega_x )=0 \, ,
\end{equation}
where $\widetilde \omega = c \widetilde k$. Next we consider
Eq.~(\ref{cw2}) for small $\Lambda\ll 1$ and assume that $\Lambda
\ll \Lambda_z, \Lambda_x$  for the solutions of our interest (this
is known to be the case modulation solutions describing dispersive
shock waves in weakly dispersive systems, where at the soliton edge
one has $k \to 0$ but $|k_x|, |k_z| \to \infty$ --- see \cite{el}
for the general discussion of this behaviour and \cite{gp74} for the
detailed calculations in the KdV case). Then to leading order we get
the characteristic equation
\begin{equation}\label{cl}
\frac{\partial \Lambda}{\partial z} + \frac{\widetilde
\Omega}{\kappa} \frac{\partial \Lambda}{\partial x}=0 \, ,
\end{equation}
which is to say
\begin{equation}\label{c2}
\Lambda= \Lambda_0  \qquad \hbox{on} \qquad
\frac{dx}{dz}=\dfrac{\widetilde \Omega (\overline \rho,
\kappa)}{\kappa}\, ,
\end{equation}
where $\Lambda_0 \ll 1$ is a constant. In particular, when
$\Lambda_0=0$ the characteristic (\ref{c2}) specifies the trailing
edge (see (\ref{kinematic})). Now, considering Eq.~(\ref{cw2}) along
the characteristic family $dx/dz=\widetilde \Omega/\kappa$ and using
$\widetilde k = \kappa$, $\widetilde \omega = \widetilde \Omega$ to
leading order, we obtain
\begin{equation}\label{cws}
\kappa_z+\widetilde \Omega_x=0 \qquad \hbox{on} \qquad
\frac{dx}{dz}=\dfrac{\widetilde \Omega (\overline \rho,
\kappa)}{\kappa}\, .
\end{equation}
We note that equation $\kappa_z+\widetilde \Omega_x=0$ arises as a
``soliton wave number'' conservation law in the traditional
perturbation theory for a single soliton (see, for instance,
\cite{grim79}) but to be consistent with the full modulation theory
it should be considered along the soliton path $dx/dz=c_{sol}=
\widetilde \Omega/\kappa$.

Since $\overline \rho$ and $\kappa$ cannot be specified
independently on one characteristic, there should exist a local
relationship $\kappa(\overline \rho)$ consistent with the system
(\ref{sw2}), (\ref{cws}). Substituting $\kappa=\kappa(\overline
\rho)$ into (\ref{cws}) and using (\ref{sw2}) we obtain
\begin{equation}\label{13-5}
    \frac{d\kappa}{d \overline\rho}=\frac{\prt\widetilde{\Omega}/\prt\overline \rho}
    {V_+-\prt\widetilde{\Omega}/\prt \kappa} .
\end{equation}
The initial condition for  the ordinary differential equation
(\ref{13-5}) follows from the requirement that the obtained
dependence $\kappa(\rho)$ should be applicable to the case of the
zero-intensity dispersive shock wave, which corresponds to initial
values $\rho^-=\rho^+=1$. In this case, the width of solitons gets
infinitely large, that is $\kappa\to0$ in the limit $\rho\to\rho^+$;
this follows also from Eq.~(\ref{7-2}) in the limit
$\rho_m\to\rho_b$.  Hence we require $\kappa(1)=0$.

According to the kinematic condition (\ref{kinematic}) the velocity
of the soliton edge is equal to the soliton velocity, so we have
\begin{equation}\label{11-3}
    s^-=\frac{dx^-}{dz}=\frac{\widetilde{\Omega}({\rho}^-,
    \kappa^-)}{\kappa^-},
\end{equation}
where  $\kappa^-= \kappa (\rho^-)$.

By introducing a new variable
\begin{equation}\label{13-7}
    \widetilde{\alpha}=\sqrt{1-\frac{\kappa^2(1+\ga \overline\rho)^2}{4 \overline\rho}}
\end{equation}
instead of $\kappa$, Eq.~(\ref{13-5}) reduces to the ordinary
differential equation
\begin{equation}\label{13-6}
    \frac{d\widetilde{\alpha}}{d \overline\rho}=-\frac{(1+\widetilde{\alpha})
    [1+3\ga \overline \rho+2\widetilde{\alpha}(1-\ga \overline\rho)]}{2\overline \rho(1+\ga \overline\rho)
    (1+2\widetilde{\alpha})}
\end{equation}
with the initial condition
\begin{equation}\label{14-1}
    \widetilde{\alpha}(1)=1.
\end{equation}
When the function $\widetilde{\alpha}(\rho)$ is found, the velocity
of the trailing soliton is determined by Eqs.~(\ref{11-3}),
(\ref{13-4}), (\ref{13-7}) as
\begin{equation}\label{14-2}
    s^-=\frac2{\sqrt{\ga}}(\arctan\sqrt{\ga\rho^-}-\arctan\sqrt{\ga})
    +\frac{\sqrt{\rho^-}}{1+\ga\rho^-}\widetilde{\alpha}(\rho^-).
\end{equation}
Then the amplitude $a=\rho^- -\rho_m$ of the trailing soliton as a
function of the intensity jump $\rho^-$ across the dispersive shock
can be found from the equation (\ref{speedamp}) with $c=s^-$,
$u_b=u^-$, $\rho_b=\rho^-$:
\begin{equation}\label{14-3}
\frac{\rho^-\widetilde{\alpha}^2(\rho^-)}{(1+\ga\rho^-)^2} =
\frac{2(\rho^- - a)}{\gamma
a}\left[\frac{1}{\gamma a}\ln \frac{1+\gamma\rho^-}{1+\gamma (\rho^-
- a)}- \frac{1}{1+\gamma \rho^-} \right].
\end{equation}

Again, in the case  $\ga=0$ corresponding to the NLS equation, all
the formulae can be written down explicitly: Eq.~(\ref{13-6})
reduces to
\begin{equation}\label{14-4}
    \frac{d\widetilde{\alpha}}{d\overline\rho}=-\frac{1+\widetilde{\alpha}}{2\overline\rho}
    \, ,
\end{equation}
and its solution satisfying the boundary condition (\ref{14-1}) is
\begin{equation}\label{14-5}
    \widetilde{\alpha}(\overline\rho)=\frac2{\sqrt{\overline\rho}}-1.
\end{equation}
Then Eqs.~(\ref{14-2}) and (\ref{14-3}) give
\begin{equation}\label{14-6}
    s^-=\sqrt{\rho^-}
\end{equation}
and
\begin{equation}\label{14-7}
    a=4(\sqrt{\rho^-}-1)
\end{equation}
respectively, in agreement with known results \cite{gk87}.
\begin{figure}[bt]
\includegraphics[width=8cm,height=6cm,clip]{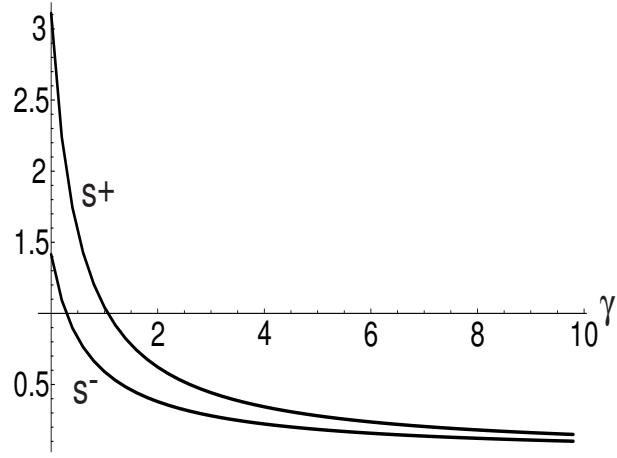}
\caption{Dependence of velocities $s^+$ and $s^-$ on the saturation
parameter $\ga$ for fixed values of the intensities at two sides of
the dispersive shock: $\rho^-=2$ and $\rho^+=1$. } \label{fig6}
\end{figure}
Again for small $\gamma$ we can find the correction to
Eq.~(\ref{14-6}) in an explicit form. If we denote
$\widetilde{\alpha}=\widetilde{\alpha}_0+ \widetilde{\alpha}_1$,
where $\widetilde{\alpha}_0$ is given by Eq.~(\ref{14-5}), then
$\widetilde{\alpha}_1$ satisfies the equation
\begin{equation}\label{12a-1}
    \frac{d\widetilde{\alpha}_1}{d\overline\rho}=-\frac{\widetilde{\alpha}_1}{2\overline\rho}
    +\frac{8/\sqrt{\overline\rho}-6}{4/\sqrt{\rho}-1}\frac{\gamma}{\sqrt{\overline\rho}},
    \qquad \widetilde{\alpha}_1(1)=0,
\end{equation}
which is readily integrated to give
\begin{equation}\label{12a-2}
\widetilde{\alpha}_1(\rho^-)=\frac{2\gamma}{\sqrt{\rho^-}}\left\{\rho^--1
+64\left[\ln\frac{4-\sqrt{\rho^-}}3+\frac{\sqrt{\rho^-}-1}4+\frac{\rho^--1}{32}
\right]\right\} \, ,
\end{equation}
and hence
\begin{equation}\label{12a-3}
    s^-=\sqrt{\rho^-}(1+\widetilde{\alpha}_1(\rho^-))-
    \left[\frac23(\rho^-\sqrt{\rho^-}-1)+\rho^-(2-\sqrt{\rho^-})\right]\gamma.
\end{equation}
It is worth noticing that this perturbation approach breaks down for
$\rho^-\geq16$ because of logarithmic divergence in
Eq.~(\ref{12a-2}) as $\rho^-\to 16-0$. The velocities of the
dispersive shock edges as functions of the saturation parameter
$\ga$ are shown in Fig~6. As we see, the presence of even small
values of the saturation parameters $\ga$ change the expansion
velocities considerably compared with the NLS case $\ga=0$ because
the saturation effects diminish the effective nonlinearity which
forces the intensive light beam to expand.

\subsubsection{Characteristic velocity ordering}
From general point of view, it is important to note that a
simple-wave dispersive shock considered above is subject to the
conditions similar to ``entropy'' conditions in viscous shocks
theory \cite{el,ekt}. Basically, these conditions require that the
number of independent parameters characterizing the modulation
solution for the dispersive shock must be equal to the number of
characteristics families transferring initial data from the $x$ axis
{\it into} the dispersive shock region in the $(x,t)$ plane. For the
photorefractive dispersive shock we have four parameters
characterizing the initial step (\ref{step}) and one algebraic
restriction due to the simple-wave transition condition (\ref{6-1}).
Thus, the number of independent parameters is three. Then, analysis
of the characteristic directions at the edges of the dispersive
shock waves leads to the following inequalities establishing the
ordering between the velocities of the dispersive shock edges and
the characteristic velocities (\ref{4-4}) of the dispersionless
system :
\begin{equation}\label{entropy}
    V^-_- <s^- < V^-_+, \quad V^+_+<s^+,\quad s^+ > s^-,
\end{equation}
where subscripts correspond to definitions (\ref{4-4}) and
superscripts to two edges of the dispersive shock with constant
values of $\rho^{\pm}$ and $u^{\pm}$. Inequalities (\ref{entropy})
provide consistency of the above analytical construction for the
derivation of the dispersive shock edges, which heavily relies on
the properties of characteristics. We have checked that inequalities
(\ref{entropy}) are satisfied for a wide range of parameters. As an
illustration, we present in Fig.~7 the plots of the characteristic
speeds in the  simple-wave photorefractive dispersive shock for $\gamma=0.2$ as
functions of the intensity jump across the shock. One can see that
the ordering (\ref{entropy}) is satisfied.
\begin{figure}[bt]
\includegraphics[width=8cm,height=6cm,clip]{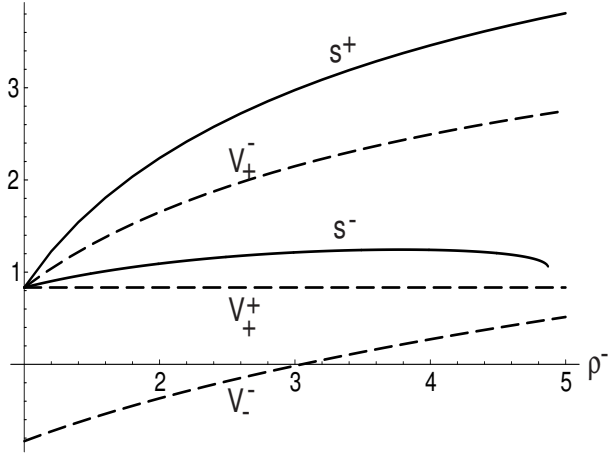}
\caption{Ordering of the characteristic velocities in the system
satisfies inequalities (\ref{entropy}). } \label{fig7}
\end{figure}

\subsubsection{Vacuum point}
\begin{figure}[bt]
\includegraphics[width=8cm,height=6cm,clip]{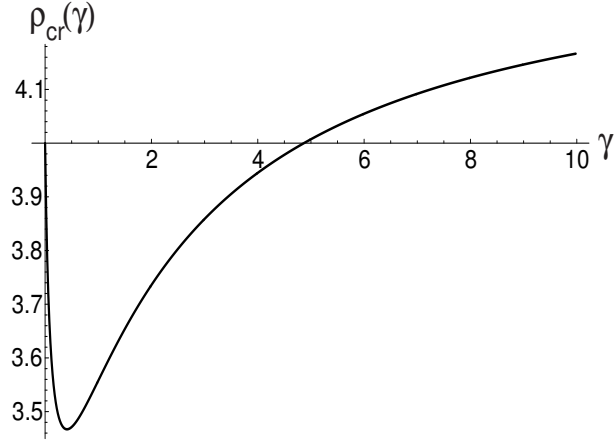}
\caption{Dependence of the critical intensity $\rho^-$ at the
trailing edge of the dispersive shock on the saturation parameter
$\gamma$ for fixed value of the intensity $\rho^+=1$ at the leading
edge. } \label{fig8}
\end{figure}
\begin{figure}[bt]
\includegraphics[width=8cm,height=6cm,clip]{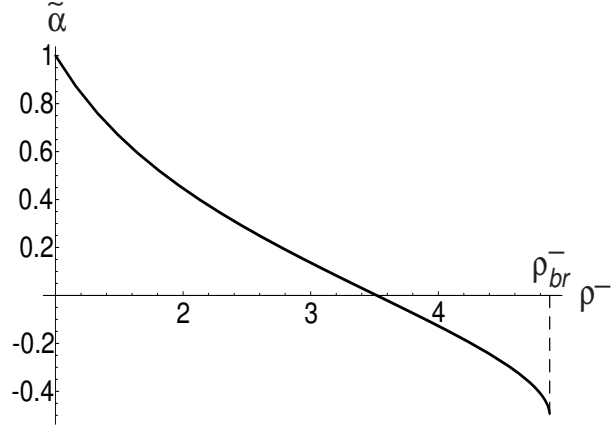}
\caption{Dependence of the variable $\widetilde{\al}$ on the
intensity $\rho^-$ at the trailing edge of the dispersive shock for
fixed value of the intensity $\rho^+=1$ at the leading edge at
$\ga=0.2$; the ``termination'' point corresponds to the intensity
$\rho^-_{br}=4.873$ where analytical theory loses its applicability.
} \label{fig9}
\end{figure}
We now investigate dependence of the main properties of the
dispersive shock wave on the value of the intensity jump across the
shock, which is equal to the value $\rho^-$ at the trailing edge as
the value $\rho^+=1$ at the leading edge is fixed (of course, we
assume $u^+=0$ and $u^-$ given by Eq.~(\ref{5-6})).

It is clear already from the simplest case $\gamma=0$ that there is
a possibility for the value $\rho_m$ at the minimum of the trailing
dark soliton to become zero (or, which is the same, $a=\rho^-$) for
a certain value of the initial jump $\rho^-$. Then it follows from
(\ref{14-7}) that this happens at $\rho^-=4$. This gives rise to a
vacuum point with $\rho=0$ at the trailing edge of the dispersive
shock \cite{eggk95}. When the initial step $\rho^->4$, the vacuum
point occurs at some $x_v$ inside the dispersive shock zone, $x^-
<x_v < x^+$, and the typical profile of the shock changes (see
\cite{eggk95}). The appearance of the vacuum point in the dispersive
shock is manifested by the singularity in the profile of $u$ at
$x=x_v$ but the ``momentum'' $\rho u$ remains finite.

For the photorefractive case, when $\gamma \ne 0$, the critical
value of $\rho^-=\rho^-_{cr}$ corresponding to the appearance of the
vacuum point at the trailing edge of the dispersive shock can be
found by putting $\rho^-=a$ in (\ref{14-3}) which immediately yields
the equation for $\rho^-_{cr}$
\begin{equation}\label{}
 \widetilde{\alpha} (\rho^-_{cr}) = 0 ,
\end{equation}
where $\tilde \alpha (\rho)$ is the solution of the ordinary
differential equation (\ref{13-6}). The dependence
$\rho^-_{cr}(\gamma)$ is shown in Fig.~8. Comparison of
Eq.~(\ref{14-2}) with Eq.~(\ref{5-6}) shows that at the critical
point $\rho^-=\rho_{cr}^-$ we have $s^-=u^-$, that is the trailing
soliton is at rest in the reference frame of the intermediate
constant state in the decay of an initial discontinuity (\ref{step})
(see Section III.A).

The dependence of $\widetilde \alpha$ on $\rho^-$ is shown in
Fig.~9. One should note that the change of sign of $\widetilde
\alpha$ at $\rho^- = \rho_{cr}$ does not constitute nonphysical
behaviour even though $\widetilde \alpha$ as defined by (\ref{13-7})
is a positive value. In fact, for $\rho^- > \rho_{cr}$, the velocity
$u$ changes its sign at $x=x_v$ so that the trailing edge of such a
``supercritical'' dispersive shock wave propagates to the left
relative to the vacuum point. To incorporate this change, one should
use another branch in the linear dispersion relation (\ref{dr})
which leads to the change of the sign in the definition of
$\widetilde \alpha$. As a result, the consistent change of signs in
(\ref{13-4}) and (\ref{13-6}) leads to the same result for the
trailing edge speed $s^-$ defined by (\ref{12a-3}).

One can also see from (\ref{13-6}) that a singularity in the
behaviour of $\widetilde \alpha (\rho^-)$ is expected at some
``termination point'' $\rho^-=\rho^-_{br}$ satisfying $2\alpha
(\rho^-)+1=0$ for $\gamma \ne 0$. For $\gamma =0.2$ the value
$\rho^-_{br}\approx 4.873$.  This singularity has also its
counterpart in the perturbation theory represented by
Eq.~(\ref{12a-2}). The described pathology in the modulation
solution for $\rho^- \ge \rho^-_{br}$, however, is not confirmed by
the direct numerical solutions (see Section IV below) and does not
seem to have physical sense. One of the explanations of such a
discrepancy is that for large values of $\rho^-$ the accepted
assumption of applicability of the single-phase modulation theory
can fail. Indeed, the developed theory is based on the supposition
that solutions of our non-integrable photo-refractive system
(\ref{dh}) behave qualitatively similar to their counterparts in the
integrable NLS equation case so that the dispersive shock wave can
be described with high accuracy by the single-phase modulated
solution. However, such a supposition can fail in the regions where
a drastic change of the behavior of a modulated wave takes place.
Just this situation occurs in the vicinity of the vacuum point, at
which the profile of $u(x)$ has a singularity. So one can expect
some discrepancy between  predictions of the modulation theory and
exact numerical solutions for the dispersive shock waves with
$\rho^-$ sufficiently close to or grater than $\rho_{cr}$. As a
rough estimate for $\rho^-_{cr}$ one can use the value
$\rho^-_{cr}=4$ obtained for the integrable NLS equation. Since, by
definition, $\widetilde \alpha (\rho^-_{br})=-1/2<0$ for all
$\gamma>0$, on can conclude that one always has $\rho^-_{br}>
\rho^-_{cr}$ so the predictions of the developed modulation theory
can become unreliable for such large intensity jumps across the
dispersive shock.

\subsection{Number of solitons generated from a localized initial
pulse}

Now we consider an asymptotic evolution of a large-scale decaying
initial disturbance
\begin{equation}\label{hole}
\rho(x,0)= \rho_0(x)\le 1 , \quad u(x,0)=u_0(x); \qquad \rho_0(x)
\to 1 \, ,\ \ u_0(x) \to 0 \quad \hbox{as} \quad |x| \to \infty ,
\end{equation}
so that the typical spatial scale of this disturbance $L \gg 1$. As
the numerical simulations for the GNLS equation show, such an
initial ``well'' generally decays as $z \to \infty$ into two groups
of dark solitons propagating in opposite directions, which is
consistent with the ``two-wave'' nature of the GNLS equation. For
$\gamma=0$ the dynamics is described by the integrable NLS equation
and the soliton parameters are found from the generalized
Bohr-Sommerfeld rule \cite{kku}. In the present non-integrable case
of the GNLS equation (\ref{dh}) these parameters can be obtained by
an extension of the modulation method of obtaining the parameters of
the dispersive shock wave for the case when the initial distribution
corresponds to the simple wave solution of the dispersionless
equations, that is one of the Riemann invariants (\ref{4-2}) is
supposed to be constant. This extension has been developed in
\cite{egs} in the context of fully nonlinear shallow-water waves and
we shall use it here to derive the formula for the total number of
solitons resulting from the initial disturbance (\ref{hole}). First,
we assume that for the {\it large-scale} initial data (\ref{hole})
one can neglect the contribution of the radiation into the
asymptotic as $z \to \infty$ solution, which implies that the whole
initial disturbance eventually transforms into solitons (this is
known to be the case for the integrable NLS equation and is also
confirmed by our numerical simulations for the GNLS equation). Next,
we notice that this transformation into solitons occurs via an
intermediate stage of the dispersive shock wave formation so we can
apply the general modulation theory to its description and then to
make some inferences pertaining to the eventual soliton train state
as $z \to \infty$.

 For definiteness, we consider here the
right-propagating dispersive shock wave forming from the profile
(\ref{hole}) satisfying an additional simple-wave restriction
(\ref{5-6})
\begin{equation}\label{sw}
u_0(x) =\frac{2}{\sqrt{\gamma}}(\arctan \sqrt{\gamma \rho_0(x)} -
\arctan \sqrt{\gamma})\,  .
\end{equation}
We consider the wave number conservation law (\ref{10-2}), which is
one of the modulation equations describing the dispersive shock
wave. For the considered case with decaying at infinity initial
profile (\ref{hole}) we have $k \to 0$ as $|x| \to \infty$ and,
therefore, equation (\ref{10-2}) implies conservation of the total
number of waves,
\begin{equation}\label{N1}
N \cong \frac{1}{2\pi}\int^{+\infty}_{-\infty} k dx =
\hbox{constant} .
\end{equation}
We use an approximate equality sign here due to asymptotic character
of the modulation theory which inherently cannot predict an integer
$N$ exactly. In the Whitham description of the dispersive shock
wave, the $x$-axis is subdivided, after the wave breaking at
$z>z_b$, into three regions described in Section III.C :
\begin{equation}\label{19-1}
    -\infty< x< x^-(z),\quad x^-(z)\leq x\leq x^+(z),\quad
    x^+(z)< x<\infty\, ,
\end{equation}
where $x^{\pm}(z)$ are the boundaries of the dispersive shock wave.
Generally, these boundaries are not straight lines as in the case of
the decay of the initial step-like pulse considered above but their
nature as characteristics of the modulation system remains
unchanged. In view of (\ref{19-1}), the integral in (\ref{N1}) can
be expressed as a sum of three integrals,
\begin{equation}\label{20-1}
    N \cong \frac{1}{2\pi}\left\{\int^{x^-(z)}_{-\infty} k(x,z) dx
    +\int^{x^+(z)}_{x^-(z)} k(x,z) dx+\int_{x^+(z)}^{\infty} k(x,z) dx
    \right\}.
\end{equation}
To apply formula (\ref{20-1}) we need first to define the wavenumber
$k$ outside the dispersive shock wave as it has been actually
defined so far only within the nonlinear modulated wave region
$[x^-(z), x^+(z)]$. The extended definition of $k$ should be
consistent with the matching conditions (\ref{lead}), (\ref{trail})
for all $z$.

We know that at the soliton edge $x^-(z)$ of the Whitham zone we
have $k(x^-(z),z)=0$, so we can safely put $k(x,z)=0$ in the region
$x< x^-(z)$ and, hence, the first integral vanishes. At the same
time, the value of $k$ is not explicitly prescribed at the leading
edge $x^+(z)$ by the boundary condition (\ref{lead}) but is rather
determined as a function of $\rho$ due to the fact that the leading
edge is a characteristic of the modulation system
--- see Section III.C. The dependence $k^+(\rho)$ is determined then by
the ordinary differential equation (\ref{12-2}) (we note that the
simple-wave transition condition (\ref{5-6}) is already embedded in
(\ref{12-2}) and is consistent with the initial conditions
(\ref{sw})). This ordinary differential equation should be, again,
solved with the initial condition $k(\rho^-)=0$, and now
$\rho^-=\rho_b\cong 1$ where we have taken into account that for
large pulse (\ref{hole}) the wave breaking occurs close to the
background intensity, $\rho_b\cong 1$. Thus, we get the
characteristic integral $k=k^+(\rho)$ along the leading edge. The
intensity  $\rho(x, z)$  in the downstream region $x> x^+(z)$
satisfies the simple-wave dispersionless equation
\begin{equation}\label{20-2}
    \rho_z+V_+(\rho)\rho_x=0
\end{equation}
with the initial condition $\rho(x,0)=\rho_0(x)$, i.e. the solution
is given implicitly by $\rho=\rho_0(x-V_+(\rho)z)$. Therefore, to be
consistent with the boundary values of $k$ prescribed by the
characteristic integral of the modulation equations, we have to
define the wave number downstream the dispersive shock wave as
$k^+(\rho(x,z))$, where $\rho(x,z)$ is the aforementioned
simple-wave solution. Then, at $z=0$ we get an effective initial
distribution of $k$ in terms of the initial data for $\rho$ given by
(\ref{hole}):
\begin{equation}\label{k0}
 k(x,0)=k^+(\rho_0(x))\,
\end{equation}
for $x \ge x_b$, where $x_b$ is the coordinate of the breaking
point; obviously $x_b=x^-(z_b)=x^+(z_b)$. Note that this definition
is also consistent with our definition $k \equiv 0$ upstream of the
dispersive shock wave, since $k^+(1)=0$. Thus, (\ref{k0}) describes
initial data for the wave number for all $x$. The function $k(x,0)$
can be interpreted as the wave number distribution for a ``virtual''
linear modulated wave which accompanies the initial hydrodynamic
distributions $(\rho(x,0)$, $u(x,0)$ and transforms, after the wave
breaking, into the dispersive shock and, eventually, into a train of
solitons.

Now, we consider (\ref{20-1}) for $z=0$ and notice that, since  the
second integral disappears for $z<z_b$, $z_b>0$ (there is no
dispersive shock before the breaking point formation so we put
$x^+(z)=x^-(z)$ for $z\le z_b$), this expression reduces to
\begin{equation}\label{20-3}
    N\cong \int_{-\infty}^{\infty} k(x,0)dx=
    \int_{-\infty}^{\infty} k^+(\rho_0(x))dx.
\end{equation}

As was shown in Section III.C, it is convenient to introduce an
auxiliary function $\alpha(\rho)$ instead of $k^+(\rho)$ according
to (\ref{12-3}) so that
\begin{equation}\label{}
 k^+=2\frac{\sqrt{\rho(\alpha^2 - 1)}}{1+\gamma \rho}.
\end{equation}
Then $\alpha(\rho)$ satisfies the ordinary differential equation
(\ref{12-4}) with the initial condition $\alpha(1)=1$.
As a result, the number of solitons as $z \to \infty$ is determined
by the formula
\begin{equation}\label{N2}
 N \cong \frac{1}{2\pi}\int^{+\infty}_{-\infty}k(x,0)dx =
 \frac{1}{\pi}\int^{+\infty}_{-\infty}
\frac{\sqrt{\rho_0(x)(\alpha_0^2(x)-1)}}{1+\gamma \rho_0(x)} dx,
\end{equation}
where $\alpha_0(x)=\alpha(\rho_0(x))$.

When $\gamma=0$, the solution $\alpha(\rho)$ of (\ref{12-4}) is
given by (\ref{13-2}) and assumes here the form
\begin{equation}\label{}
 \alpha=\frac{2}{\sqrt{\rho}}-1\ \qquad \hbox{for} \quad \gamma=0 .
\end{equation}
Then, for the total number of solitons we have from (\ref{N2})
\begin{equation}\label{}
   N \cong \frac{1}{\pi}\int^{+\infty}_{-\infty}\sqrt{(2-\sqrt{\rho_0(x)})^2 -
\rho_0(x)}=\frac{2}{\pi}
\int^{+\infty}_{-\infty}\sqrt{1-\rho_0^{1/2}}dx \, \qquad \hbox{for}
\quad \gamma=0,
\end{equation}
which agrees with the ``simple-wave'' reduction of the semi-classical
quantization results for the defocusing NLS equation obtained in
\cite{kku}.

\section{Numerical simulations of nonlinear waves in
photorefractive media}

In this Section, we compare the analytical predictions of the
preceding Sections  with the results of direct numerical simulation
of the formation of dispersive shock waves in photorefractive
equation (\ref{1-5}).

First, we have studied numerically evolution of the step-like pulse.
The corresponding results are shown in Fig.~10. As we see, all the
parameters (velocities of the edges of the rarefaction wave and the
dispersive shock, intensity of the intermediate state) are in a good
agreement with the analytical predictions of Section~III.A.
\begin{figure}[bt]
\includegraphics[width=9cm,height=7cm,clip]{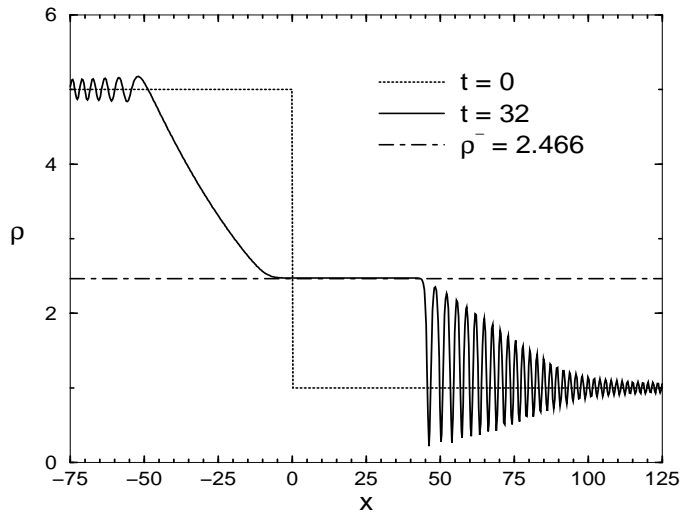}
\caption{Evolution of the initial step-like pulse with $\rho_0=5$
and $\rho=1$ for the case of $\ga=0.1$. The general structure
confirms formation of a rarefaction wave, a dispersive shock and an
intermediate constant state in between. Intensity $\rho^-=2.466$
calculated according to Eq.~(\ref{5-5}) coincides with the numerical
result for the intensity of the intermediate state. Coordinates of
the edges of the rarefaction wave at $t=32$ calculated analytically
are equal to $x_1^-=-47.7$, $x_1^+=-9.02$ for the rarefaction wave
and $x_2^-=42.57$, $x_2^+=99.52$. One can see that they agree quite
well with numerical results. Small-amplitude waves generated at
around $x=-50$ correspond to the linear dispersive ``resolution'' of
the weak discontinuity occurring at the trailing edge of the
rarefaction wave.} \label{fig10}
\end{figure}

We have constructed the dependence of $\rho^-$ and $u^-$ on the
saturation parameter $\ga$ using the results of the numerical
simulations. The results shown in Fig.~11 agree very well with the
analytical predictions based on the ``simple-wave'' jump condition
(\ref{6-1}) which is applicable for not too large values of $\rho^-$
($ \lesssim 4$) such that the vacuum point is not formed.
\begin{figure}[bt]
\includegraphics[width=8cm,height=6cm,clip]{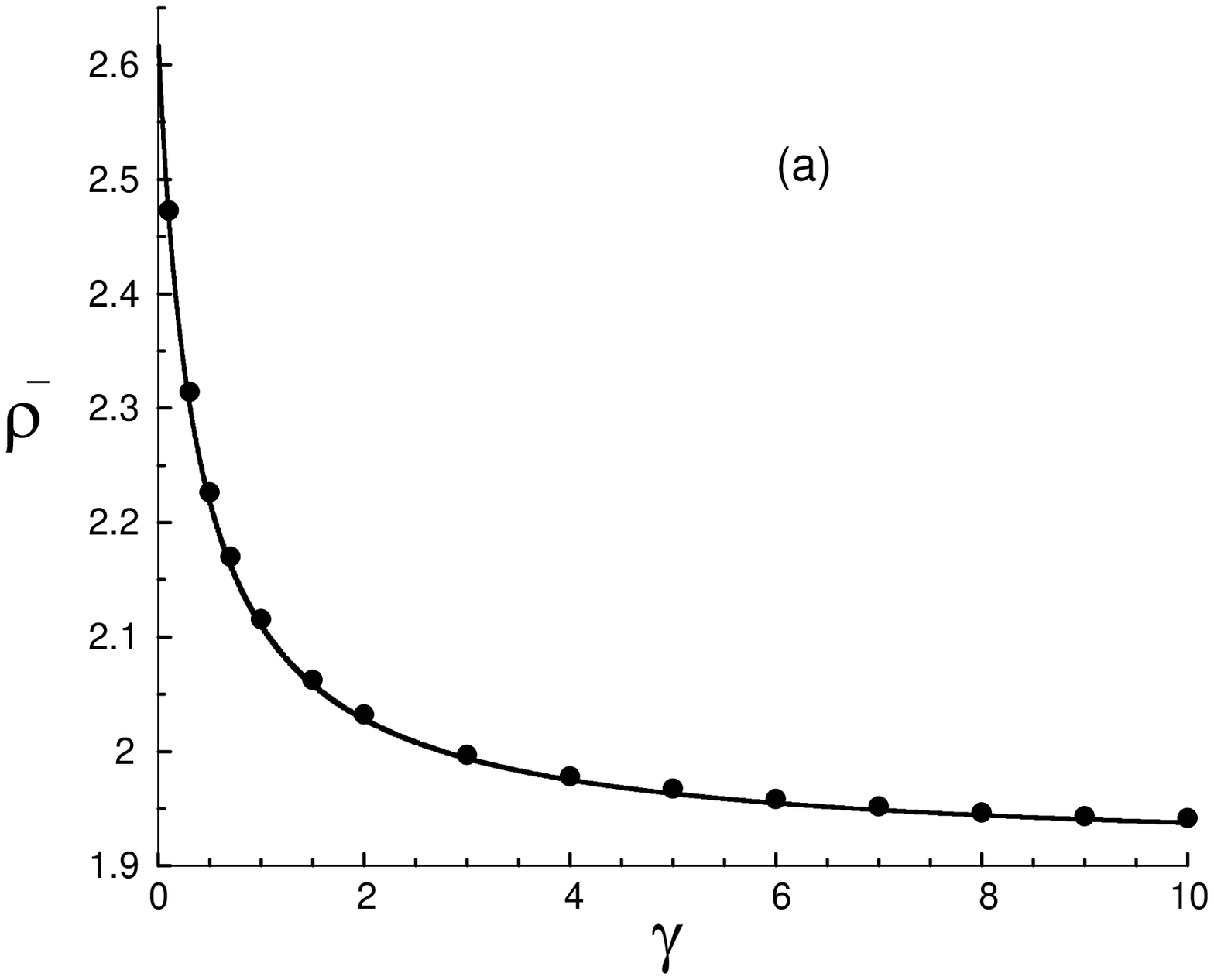}
\hspace{1cm}
\includegraphics[width=8cm,height=6cm,clip]{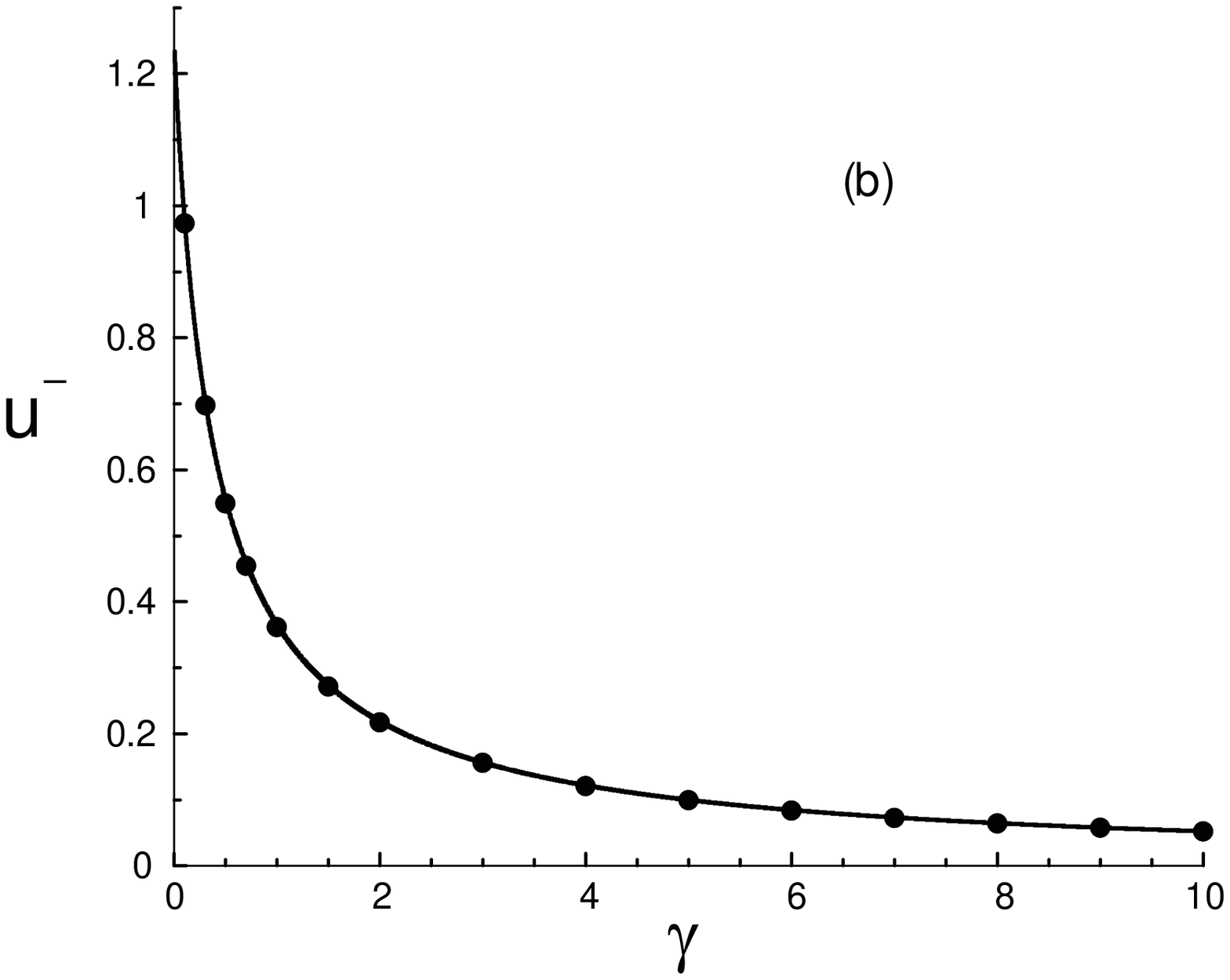}
\caption{Dependence of intermediate values (solid lines) of
intensity (a) and transverse wave vector (b) on the saturation
parameter $\gamma$ for fixed values of the initial discontinuity
parameters: $\rho_0=5$, $u_0=0$ for $x<0$ and $\rho^+=1$, $u^+=0$
for $x>0$ at $z=0$. Numerically calculated values are shown by
crosses. } \label{fig11}
\end{figure}
In Fig.~12 we show the dependence of the edge ``velocities''
$s^{\pm}$ on the intermediate intensity $\rho^-$ (with $u^-$
calculated according to ``simple-wave'' jump condition (\ref{5-6})).
As we see, a good agreement is observed for $\rho^-<4$.
\begin{figure}[bt]
\includegraphics[width=8cm,height=5cm,clip]{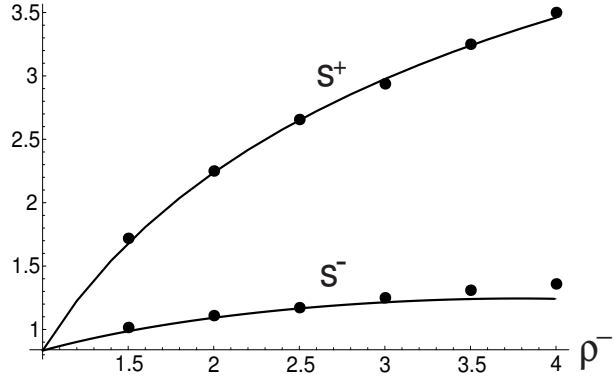}
\caption{Dependence of $s^{\pm}$ on $\rho^-$ (with $u^-$
calculated according to (\ref{5-6})); $\gamma=0.2$.
Solid lines correspond to
analytical formulae (\ref{12-7}) and (\ref{14-2}), and dots
correspond to results of numeric simulations.
} \label{fig12}
\end{figure}
However, as $\rho^-$ increases with growth of $\rho_0$ and becomes
greater than $\rho^-_{cr}\simeq 4$, Eq.~(\ref{5-6}) no longer yields
the values of $u^-$ compatible with the prescribed value of $\rho^-$
so that only a single right-propagating dispersive shock is
generated; this is illustrated by Fig.~13, where a new
``intermediate'' region of constant flow is seen to be formed which
matches with the dispersive shock propagating to the right, while
another dispersive shock is apparently forming to the left of this
new constant state providing matching with $\rho_0$.
\begin{figure}[bt]
\includegraphics[width=8cm,height=6cm,clip]{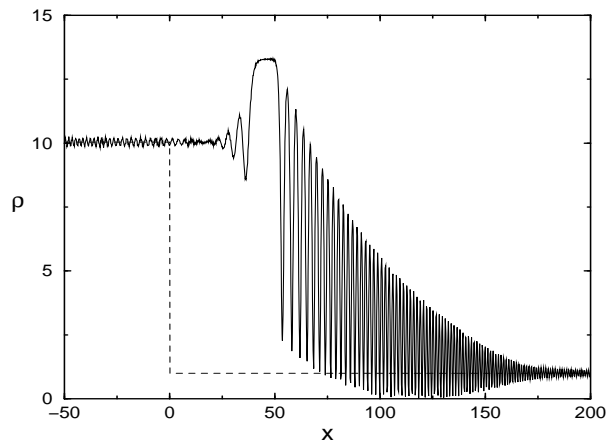}
\caption{Dispersive shock evolving from the step-like pulse with
$\rho^-$ and $u^-$ related by the ``simple-wave'' jump condition for
large value of $\rho^-=10$ much greater than $\rho^-=4$. Occurrence
of a vacuum point in the region between $x=100$ and $x=150$ is
clearly seen. A new intermediate constant state is formed in the
region behind the dispersive shock showing that the simple-wave jump
condition (\ref{5-6}) does not prevent anymore the formation of the
second wave for large values of $\rho^-$.} \label{fig13}
\end{figure}
Surprisingly, we have found that the large-amplitude dispersive
shock wave transition between the new intermediate constant state
and $\rho=1$ now satisfies a classical shock jump condition which
follows from the balance of ``mass'' and ``momentum'' across the
shock as it takes place in classical dissipative shocks. Using the
dispersionless equations (\ref{4-1}) represented in a conservative
form we find that formal shock jump conditions yield the dependence
\begin{equation}\label{classical}
    u^-=\frac{\sqrt{2}(\rho^--1)}{\sqrt{(\rho^-+1)(1+\gamma\rho)(1+\gamma)}}
    \, .
\end{equation}
We have checked that dependence (\ref{classical}) is indeed
satisfied very well for $\rho^->4$. The physical mechanism
supporting the appearance of the classical shock conditions in a
dissipationless system such as (\ref{dh}) is not quite clear at the
moment. We note that similar effect of appearance of the classical
shock jump condition across the expanding dispersive shock have been
recently observed in \cite{egs2006} for large-amplitude
shallow-water undular bores modeled by the  Green-Naghdi system,
which is also not integrable by the IST. At the same time, it is
known very well that for the dispersive shocks described by the
integrable NLS equation, the simple-wave jump condition is satisfied
exactly for all values of initial density jump  -- this follows from
the full modulation solution (see \cite{eggk95}, \cite{kodama},
\cite{hoefer}) and is also confirmed by our numerical simulations.
So it is possible that the described phenomenon of the appearance of
the classical shock conditions constitutes a specific manifestation
of nonintegrability in dispersive dissipationless systems which is
yet to be explored.

Next, we have compared the analytical predictions of subsection
III.D for number of dark solitons generated from a hole-like
disturbance with numerical simulations. We took the initial
distribution of intensity
\begin{equation}\label{int-N}
    \rho_0(x)=\left(1-\frac1{\cosh(0.2 x)}\right)^2
\end{equation}
and the initial distribution of transverse wave number was
calculated according to the equation (\ref{sw}). Evolution of such a
pulse according to the photorefractive equation with $\gamma=0.2$ is
illustrated by Fig.~14 where the profile of intensity is shown at
$z=100$.
\begin{figure}[bt]
\includegraphics[width=8cm,height=6cm,clip]{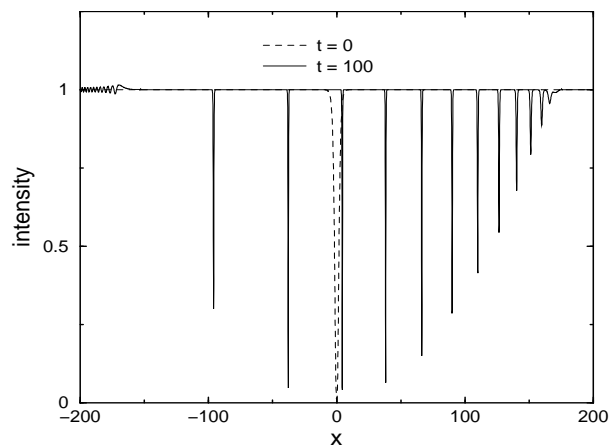}
\caption{Profile of intensity at $z=100$ evolved from the initial pulse
(\ref{int-N}) (dashed line) with initial profile of $u(x)$
calculated according to (\ref{sw}).
} \label{fig14}
\end{figure}
As we see, this pulse, after the wave breaking and formation of a
dispersive shock, evolves eventually into a number of dark solitons.
We note that the appearance of several solitons propagating to the
left does not contradict to the unidirectional restriction
guaranteed by the simple-wave initial conditions (\ref{int-N}),
(\ref{sw}) -- these left-propagating solitons occur due to
relatively high amplitude of the initial disturbance (\ref{int-N}),
which leads to the appearance of the vacuum point at the
intermediate stage of the dispersive shock wave and, therefore, to
the formation of some number of left-propagating solitons -- see
Section III.C. The  total number of created solitons calculated by
means of the modulation formula (\ref{N2}) as a function of the
saturation parameter $\gamma$ is shown by solid line in Fig.~15 and
the corresponding results of numerical simulations are indicated by
dots. Taking into account the asymptotic nature of the developed
analytical theory for this integer-valued function, and the fact
that considered initial data (\ref{int-N}) produce a vacuum point
(i.e. at some stage of the dispersive shock development the
``instantaneous'' initial jump $\rho^-> \rho^-_{cr}$), the agreement
can be considered as quite good.
\begin{figure}[bt]
\includegraphics[width=8cm,height=6cm,clip]{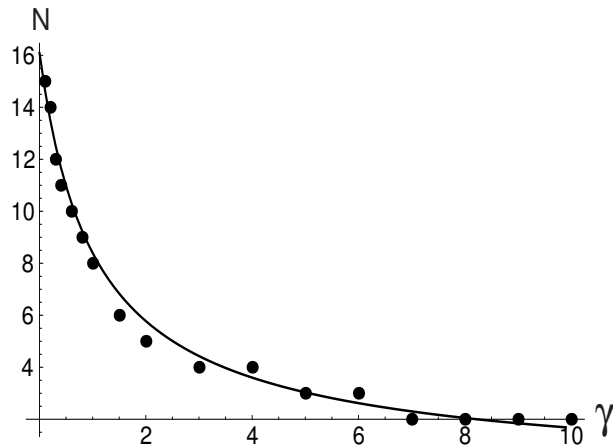}
\caption{Number of solitons $N$ as a function of $\gamma$;
solid line represents analytical dependence (\ref{N2}) and
dots correspond to numerical simulations.
} \label{fig15}
\end{figure}

In Refs.~\cite{hoefer, fleischer} the theory of dispersive shocks,
evolving from a step-like pulse according to the NLS equation
(\ref{1-6}) ($\gamma=0$) was used for qualitative explanation of
dispersive shocks with other geometries in concrete physical
situations (see also \cite{kgk04} where the NLS theory of the wave
breaking was also used for description of dispersive shocks in
Bose-Einstein condensates). In a similar way, the developed here
theory of dispersive shocks in photorefractive media can be used for
the description of experiments on generation of optical shocks. Such
experiments were described in \cite{french,fleischer} and here we
present some results based on the numerical solutions of the
photorefractive equation (\ref{1-5}) with the initial conditions
similar to the initial light distributions in the mentioned
experimental works (similar results of numerical simulation were
presented in \cite{fleischer}).
\begin{figure}[bt]
\includegraphics[width=8cm,height=6cm,clip]{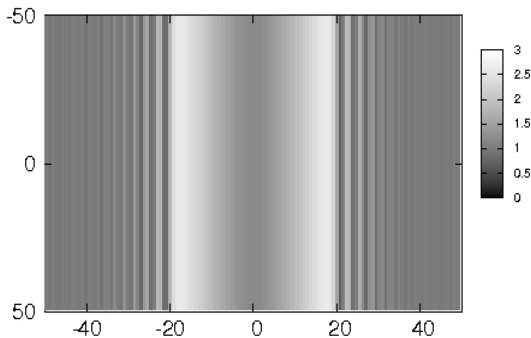}
\caption{Density plot of the output intensity evolved from a
strip-like initial distribution. } \label{fig16}
\end{figure}
\begin{figure}[bt]
\includegraphics[width=7cm,height=7cm,clip]{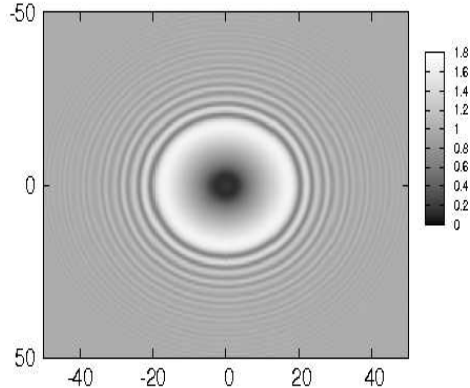}
\caption{Density plot of the output intensity evolved from a circle
initial distribution. } \label{fig17}
\end{figure}
\begin{figure}[bt]
\includegraphics[width=9cm,height=6cm,clip]{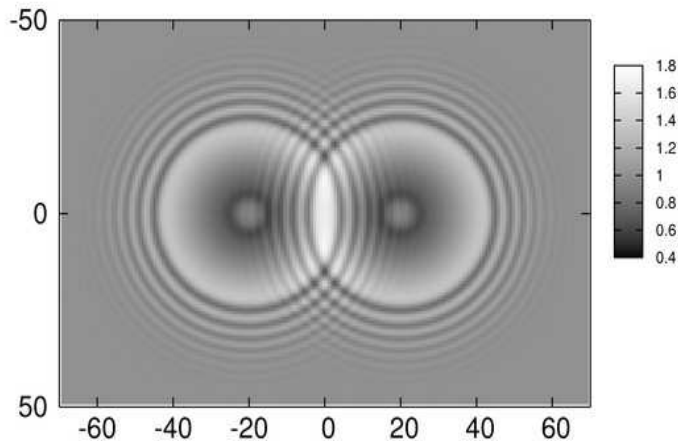}
\caption{Interaction of two circular dispersive shocks. }
\label{fig18}
\end{figure}

In \cite{fleischer} the distribution of output intensities are
presented for initial distributions in the form of a strip, a
circle, and two separated circles. We have performed numerical
simulations with similar initial conditions. In Fig.~16 we present a
density plot of the output intensity evolved, according to the
photorefractive equation with $\gamma=0.1$, from the strip-like
initial distribution given by the formula
\begin{equation}\label{init-strip}
    \rho(x)=\left\{
    \begin{array}{ll}
    1+ 5(1-x^2/25)^{0.2} \quad &\text{for}\quad |x|<5,\\
    1 \quad &\text{for}\quad |x|>5 ,
    \end{array}
    \right.
\end{equation}
which approximates with a good enough accuracy the constant values
of intensities inside the strip and outside it. Similar density plot
for the circle initial distribution is shown in Fig.~17.

As we see, in both cases the initial ``hump'' breaks with formation
of dispersive shocks---in the strip-like geometry we get two shocks
propagating in opposite directions and in circular geometry we have
a ring-like dispersive shock expanding in radial direction.

In Fig.~18 an interaction of two circular dispersive shock waves is
shown. It is remarkable that even in this two-dimensional
nonintegrable photorefractive case, the nonlinear dispersive shock
waves interact apparently elastically, without production of other
waves in the region of their overlap. It is this kind of picture
that is expected in the system with $\gamma=0$ described by the
integrable NLS equation where the interaction of two dispersive
shocks leads to the formation of a two-phase modulated wave region
described by the corresponding multiphased-averaged modulation
system \cite{bk}. While the analytical description of multiphase
nonlinear waves in photorefractive equation (\ref{dh}) is not
available, the qualitative similarity between the solution behaviour
for the nonintegrable photorefractive equation and for the NLS
equation for moderate values of initial amplitudes can be considered
as a confirmation of robustness of the modulated travelling wave
ansatz in the description of dispersive shock waves in nonintegrable
systems, at least for some reasonable range of initial amplitudes.

\section{Conclusion}

In this paper, we have developed the theory of formation of
dispersive shocks in propagation of intensive light beams in
photorefractive optical systems. The theory is based on Whitham's
modulation approach in which a dispersive shock is described as a
modulated nonlinear periodic wave and slow evolution along the
propagation axis is governed by the averaged modulation equation. In
spite of the absence of complete integrability of the equation
describing propagation of light beams in photorefractive media, the
main characteristic parameters of shocks can be determined by means
of the approach developed in \cite{el}--\cite{ekt} and based on the
study of reductions of Whitham equations for the wave regimes
realized at the boundaries of the dispersive shock. In particular,
``velocities'' of the dispersive shock edges are found as functions
of the jump of intensity across the shock as well as amplitude of
the soliton at the rear edge of the shock. The number of solitons
produced from a finite initial disturbance is also determined
analytically for initial distributions related by a so-called simple
wave condition. The analytical theory agrees very well with
numerical simulations as long as there is no vacuum point in the
shock. Appearance of a vacuum point leads to the formation of a
singularity in a ``transverse'' wavevector distribution and such a
drastic change in the wave behavior cannot be traced by the
developed approach. However, this situation occurs at very high
input intensities of a light beam so that for practical purposes the
developed theory provides accurate enough approximation.

Although the theory is essentially one-dimensional (i.e. with one
transverse space coordinate) it can give qualitative explanation of
experiments with other geometries, which is illustrated by the
results of numerical simulations.

\subsection*{Acknowledgments}

This work was supported by FAPESP (Brazil) and EPSRC (UK).
AMK thanks also RFBR (grant 05-02-17351) for partial support.

\end{document}